\documentclass[fleqn,usenatbib]{mnras}

\usepackage{newtxtext,newtxmath}

\usepackage[T1]{fontenc}
\usepackage{xcolor}
\DeclareRobustCommand{\VAN}[3]{#2}
\let\VANthebibliography\thebibliography
\def\thebibliography{\DeclareRobustCommand{\VAN}[3]{##3}\VANthebibliography}

\usepackage{graphicx}	
\usepackage{amsmath}	

\title[Atmosphere Loss in Oblique Super-Earth Collisions]{Atmosphere Loss in Oblique Super-Earth Collisions}

\author[T. R. Denman et al.]{Thomas R. Denman$^{1}$\thanks{E-mail: tom.denman@bristol.ac.uk},
Zo\"{e} M. Leinhardt$^{1}$ and  
Philip J. Carter$^{1,2}$ 
\smallskip
\\
$^{1}$School of Physics, H.H. Wills Physics Laboratory, University of Bristol, Bristol BS8 1TL, UK\\
$^{2}$Department of Earth and Planetary Sciences, University of California, Davis, One Shields Avenue, Davis, CA, 95616, USA\\
}
\date{Accepted XXX. Received YYY; in original form 2021 ???}

\pubyear{2021}

\begin{document}
\label{firstpage}
\pagerange{\pageref{firstpage}--\pageref{lastpage}}
\maketitle

\begin{abstract}

Using smoothed particle hydrodynamics we model giant impacts of Super-Earth mass rocky planets between an atmosphere-less projectile and an atmosphere-rich target. In this work we present results from head-on to grazing collisions. 
The results of the simulations fall into two broad categories: 1) one main post-collision remnant containing material from target and projectile; 2) two main post-collision remnants resulting from `erosive hit-and-run' collisions. All collisions removed at least some of the target atmosphere, in contrast to the idealised hit-and-run definition in which the target mass is unchanged.    
We find that the boundary between `hit-and-run' collisions and collisions that result in the projectile and target accreting/merging to be strongly correlated with the mutual escape velocity at the predicted point of closest approach.
Our work shows that it is very unlikely for a single giant impact to remove all of the atmosphere. For all the atmosphere to be removed, head-on impacts require roughly the energy of catastrophic disruption (i.e. permanent ejection of half the total system mass) and result in significant erosion of the mantle. We show that higher impact angle collisions, which are more common, are less efficient at atmosphere removal than head-on collisions. Therefore, single collisions that remove all the atmosphere without substantially disrupting the planet are not expected during planet formation.

\end{abstract}




\section{Introduction}

Both observations and numerical simulations indicate that planet formation often results in multiple planet systems. 
Gravitational interactions with the proto-planetary disc cause planets embedded within to migrate. This migration is typically inward and is dependent on planet mass \citep{Ward1986a}. In systems with multiple Super-Earth mass planets this process forms resonant chains at the inner edge of the disc. The large rocky cores of these Super-Earths can accrete significant primordial atmospheres from the proto-planetary disc \citep{Rogers2011}. 

When the proto-planetary disc dissipates it ceases to provide the drag force that circularises the orbits of the planets. Small orbital perturbations can destabilise any resonant chain of close orbiting Super-Earths that may have formed within the disc \citep{Barnes2004} as they are often already on the borders of stability \citep{Fang2013}. Eventually, the instability will lead to collisions between the planets or gravitational ejection of a planet from the system. As such, the systems we observe with Super-Earth mass planets typically only have a few planets. 
Observed Super-Earths in multiple planet systems are often detected orbiting close to, but not in, mean motion resonance \citep{Fabrycky2014}. Very rarely we observe a system of multiple planets in a meta-stable resonant chain (for example Trappist-1, \citealt{Gillon2017}, and Kepler-223, \citealt{Mills2016}). \citet{Volk2015} suggest such resonant chains of close orbiting planets are common in planetary formation and potentially even occurred in the early stages of our own solar system's formation but that over long time periods they become unstable leading to either destruction or collisional consolidation.

Giant impacts are not only thought to be an important formation mechanism for Super-Earths. The dissipation of the proto-planetary disc can result in orbit destabilisation for objects from planetesimals all the way up to giant planets, leading to orbital crossings and collisions. A similar resonant chain breaking scenario to the Super-Earth formation process is thought to be a possible formation scenario for Hot Jupiter systems \citep{Raymond2020}. Giant impacts are thought to be a common process in planet formation, and may allow us to observe ongoing planet formation in young planetary systems \citep[e.g.][]{Watt2021}. Our own solar system has multiple examples of planets for which giant collisions are the most probable formation mechanism, for example there is strong evidence for a giant impact having formed the Earth-Moon system \citep[e.g.][]{Hartmann1975,Asphaug2014,Lock2018} and also for a giant collision causing Uranus' unusual axis of rotation and magnetic field \citep{Kegerreis2018}. 

Another impetus for the study of giant collisions between Super-Earth sized objects is the large amount of diversity in their measured densities, both overall and between Super-Earths observed in the same system. This density diversity as a whole cannot solely be accounted for by other methods, such as XUV radiation from the central star. XUV  erodes the less dense outer layers of close orbiting planets. XUV radiation should lead to a simple orbital distance density relation due to the reduction in XUV flux with orbital distance \citep{Inamdar2015}. 
XUV radiation can explain differences in density between planets orbiting different stars but not a large amount of density difference within a multiple planet system, especially when the density does not correlate with orbital distance \citep[e.g.][]{Bonomo2019}.

\citet{Inamdar2016} shows theoretically that giant collisions are a potential explanation to the observed  density diversity because collisions can cause a significant percentage of a planet's atmosphere to be ejected. In addition, \citet{Bonomo2019}, present observational evidence for a giant impact in the Kepler 107 system. In this system planet Kepler-107b is less dense ($\rho=5.3\,\mathrm{g\,cm^{-3}}$) than its neighbour Kepler-107c orbiting externally to it  ($\rho=12.6\,\mathrm{g\,cm^{-3}}$). The K-107 system exhibits exactly the opposite situation as we would expect to occur due to XUV, which has a stronger effect on planets closer to the central star. However, as shown in the paper a large collision could have stripped the lighter material from Kepler-107c, increasing its density. 

Our research in this paper involves modelling such giant impacts between Super-Earth mass exoplanets. In particular we focus on collisions involving a rocky target which has accreted a thick hydrogen atmosphere from the proto-planetary disc and an atmosphere-less rocky projectile. 

\subsection{Previous Work}

Due to the difference in density between mantle and atmosphere, and because it is preferable in SPH for (neighbouring) particles to have similar masses most previous numerical work on giant impacts have modelled the targets without atmospheres. As an approximation this works well for smaller terrestrial planets with masses $\lesssim1\mathrm{M_\oplus}$ which typically have low atmosphere mass fractions (of the order of $1\%$ or less). 
However, \citet{Kegerreis2020} modelled these thin atmospheres, which required simulations with resolutions at the upper end of what is currently typically feasible at order $10^7$ particles.

The assumption of a thin atmosphere is not accurate for Super-Earth mass objects, however, which can often accrete large amounts of gas from the proto-planetary disk. Fortunately these thicker atmospheres do not require as high a total simulation resolution to model, as the higher atmosphere fraction means a greater percentage of the total particles in the simulation are in the atmosphere.

Whilst not going so far as to include atmosphere, \citet{Gabriel2020} investigate the effects of density stratification in planetary bodies on collisions involving Earth mass objects, i.e the effect of a planet with an internal density that decreases with radius from its centre, either due to pressure or composition. They look at the effects of this stratification in objects that are comprised of a combination of iron, silicate, and water. They show that density stratification can lead to hit-and-run style collisions at lower impact angles than constant bulk density models predict. The densities of some Super-Earths analysed in our paper are even more highly stratified than those covered in \citeauthor{Gabriel2020}, in part due to the higher masses involved and also due to the density contrast between the core and mantle material and the atmosphere.

Whilst we make comparisons to  \citet{Gabriel2020}, this paper does not follow the same method; they use raw relative kinetic energy, $K$, and its relationship to the binding energy of material in the system, as suggested by \citet{Movshovitz2016}. In this paper, we instead use the specific relative kinetic energy measure of \citet{Leinhardt2012}, which is given by:
\begin{equation}
    Q_{\mathrm{R}}=\frac{1}{2}\mu\frac{v^2}{M_{\mathrm{tot} }},
    \label{eq:Q_R}
\end{equation}
where $M_{\mathrm{tot}}$ is the total mass in the system, $\mu$ the reduced mass, and $v$ the impact velocity.
Specific energy is commonly used to construct collision scaling laws \citep[e.g.][]{Benz1999,Stewart2009,Leinhardt2012}. 
This choice of $Q_{\mathrm{R}}$ is consistent with our previous work
\citep{Denman2020},  which examined solely head-on collisions involving similar mass bodies to those covered in this paper. This paper expands on our previous work by examining collisions at many different impact parameters.

\subsection{Collision Outcomes}

The head-on impacts from our previous work \citep{Denman2020} resulted in either the majority of the projectile merging with the target, or, at higher energies, one or both being disrupted. Real collisions are unlikely to be head-on, in general we would expect an impact angle closer to $45^{\circ}$ \citep{Shoemaker1962}.

At these higher impact angles additional impact outcomes can occur. For example, if a collision is sufficiently glancing a hit-and-run can occur in which the projectile bounces off the target (and in the idealised definition does not erode it \citealt{Leinhardt2012}). Another possible impact outcome is graze-and-merge  \citep{Genda2012}, these begin similarly to hit-and-run collisions but the collision removes sufficient kinetic energy from the projectile that it can not escape the gravitational influence of the target and the projectile and target eventually merge.

\section{Methods}

Here we provide a brief summary of the numerical methods used, for more details see \citet{Denman2020}.

\subsection{Numerical code}
The simulations in this paper were carried out using the same build of the SPH code GADGET-2 \citep{Springel2005} as used in \citet{Denman2020} (code available from \citealt{Gadget2Planetary2022}). This version of GADGET-2 has been modified so it can use tabulated equations of state to model planets \citep{Marcus2009,Matija2012}. GADGET-2 was run in 'Newtonian mode' with timestep synchronisation and the standard relative cell-opening criterion. We use the standard timestep criterion described in \citet{Springel2005}, with a Courant factor of $0.1$. We also use the standard artificial viscosity formulation with a strength parameter of $0.8$.

\subsection{Initial Conditions}

Planets were modelled as either two or three component objects, with an iron core, a forsterite mantle, and a hydrogen atmosphere (target only). For all planets we built an iron core surrounded by a forsterite mantle of double the mass of the core. This choice of Earth-like composition is consistent with many previous Super-Earth studies including \citet{Marcus2009,Liu2015} and \citet{Hwang2017b}. We ran preliminary simulations to equilibrate each planet in isolation to ensure they were stable, as in \citet{Marcus2009}. For the target object we then added a hydrogen atmosphere as an outer layer and re-equilibrated as per \citet{Denman2020}. The mass of this atmosphere, $M_{\mathrm{atmos}}$, was determined via the following relation:
\begin{equation}
    \frac{M_{\mathrm{atmos}}}{M_{\oplus}}=0.01\times\left(\frac{M_{\mathrm{c\&m}}}{M_{\oplus}}\right)^3,
    \label{eq:BernAtmosPred}
\end{equation}
where $M_{\mathrm{c\&m}}$ is the combined mass of the core and mantle. This rule was determined empirically from the results of Bern model population synthesis simulations  \citep{Alibert2005,Mordasini2018} for planets $\sim10\,\mathrm{Myr}$ old (a sufficiently early time that planetary systems should still be dynamically active and thus collisions should still be occurring), and is the same as was used in \citet{Denman2020}. 

All collisions used the same target, which had a total mass of $6.25\,\mathrm{M_{\oplus}}$. This was the same mass as the intermediate mass target from \citet{Denman2020}, and had a significant atmosphere. For computational run-time reasons, using only one target meant we could test collisions at a larger range of velocities and impact angles. This target planet consisted of a $1.67\,\mathrm{M_{\oplus}} $ iron core, $3.33\,\mathrm{M_{\oplus}} $ forsterite mantle and $1.25\,\mathrm{M_{\oplus}}$ hydrogen atmosphere. We used 5 different projectiles evenly spaced in mass from $1\,\mathrm{M_{\oplus}}$ to $5\,\mathrm{M_{\oplus}}$. We did not implement atmospheres in the smaller projectiles since, due to the preferability in SPH of equal mass particles, we were not simulating at a high enough resolution for the smaller atmospheres to be well resolved. We kept the larger mass projectiles without atmosphere for ease of comparison. Target and projectiles had a $1:2$ ratio by mass of iron core to forsterite mantle with equal particle masses.

We used tabulated ANEOS/MANEOS equations of state (from \citealt{Melosh1989}) to describe the iron and forsterite (full tables are available from \citealt{GADGET2EOS2019}). The hydrogen atmospheres on the other hand were modelled using the ideal gas equation of state (built in to GADGET-2)  for simplicity.
Initial radial density and temperature profiles for the core and mantle were taken from \citet{Valencia2006}, and we used the same initial profiles for the atmosphere as in \citet{Denman2020}. Planets were constructed buy splitting each planet into radial shells and placing a number of particles at random positions in each shell proportional to the density at the shell's radius. The temperature profiles were used for initial estimates before equilibration simulations were run. During equilibration particles were forced to follow isentropes, and particle velocities were damped by a restitution factor of $50\%$ each timestep. A planet was considered settled if both the mean and maximum distance of the particles of each material from the planet centre of mass changed by less than $10^{-3}\,\mathrm{R_{\oplus}}$ between output snapshots (5000\,s).

We used the same equilibration times as given in \citeauthor{Denman2020}, $4 \times 10^5 \mathrm{s}$ for preliminary core equilibration simulations and twice this for equilibrating the atmosphere on top. Core and mantle specific entropies were set to $1.3\,\mathrm{kJ\,K^{-1}\,kg^{-1}}$  and $3.2\,\mathrm{kJ\,K^{-1}\,kg^{-1}}$ respectively. Unlike in our previous work, here the initial value of the atmosphere's pseudo-entropy was set to the higher value of $1.3\times10^{11}\mathrm{Ba\,g^{-\gamma}\,cm^{3\gamma}}$ (where $\gamma=\frac{5}{3}$ was the value of the adiabatic index). This  higher pseudo-entropy meant a more extended atmosphere than \citet{Denman2020} which better matched the initial radial density profile after equilibration was completed.

Most simulations had a resolution of $10^5$ particles in the target core and mantle as this allowed us to run a large number of them. We also ran a smaller group of simulations at both higher and lower resolutions, these typically showed good agreement with the equivalent intermediate resolution results that our analysis is based on. The standard deviation of resolution test results for total mass of the largest remnant was  $2\,\%$ the result for our standard resolution, whereas atmosphere mass in the largest remnant had a standard deviation of $7\,\%$  (a more in depth discussion of this is given in appendix \ref{ap:ResolutionTesting}). Collisions were run using the University of Bristol's phase 3 and phase 4 Bluecrystal supercomputers. Phase 3 nodes were 16 core 2.6 GHz SandyBridge processors \citep{BlueCrystal3}, whereas phase 4 nodes have two 14 core 2.4 GHz Intel E5-2680 v4 (Broadwell) CPUs \citep{BlueCrystal4}. For a typical collision simulation we used all processor cores in a single compute node; with this configuration our standard resolution runs took $\sim 7\,\mathrm{hrs}$.

\subsection{Run Parameters}
\begin{figure}
 \includegraphics[width=\columnwidth]{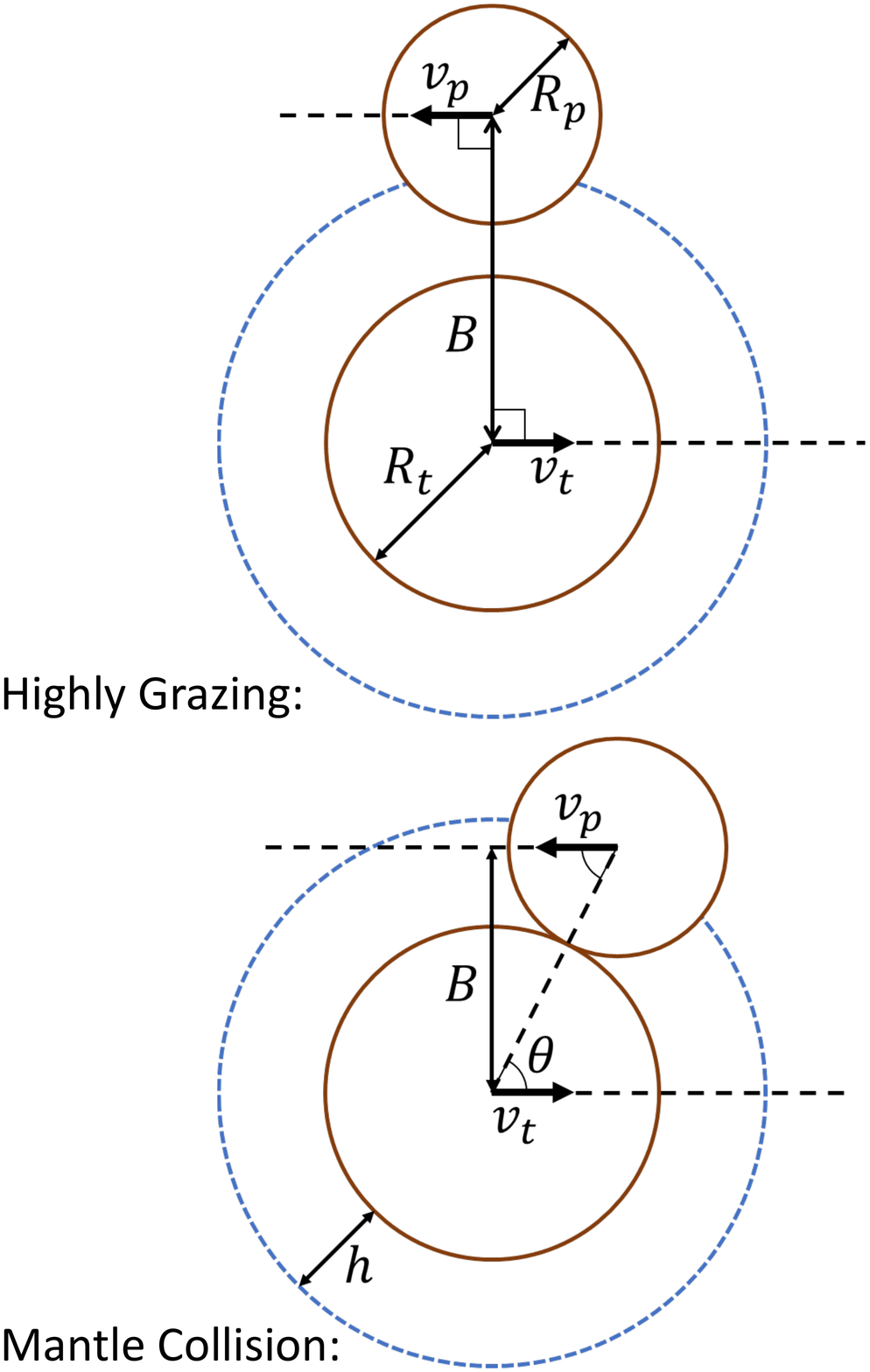}
 \caption{
    Collision geometry showing the parameters used to describe both highly grazing impacts in which the projectile only passes through the atmosphere (top) and mantle collisions where the mantles collide (bottom). A normalised impact parameter is used, $b=B/(R_{\mathrm{p}}+R_{\mathrm{t}})$. The relative velocity of an impact is $v=v_{\mathrm{t}}-v_{\mathrm{p}}$. The brown solid lines indicate the surface of the mantle of the projectile and target and the blue dotted line represents one scale height within the target's atmosphere. 
	}
 \label{fig:NM_CollisionGeometry}
\end{figure}

The simulations were run for about four times the gravitational dynamical time, $t_{\mathrm{dyn}}$, to allow enough time for gravitational resettling of material after the collision. This interaction time (one dynamical time) is given by:
\begin{equation}
    t_{\mathrm{dyn}}=\sqrt{\frac{R^3}{GM_{\mathrm{tot}}}},
    \label{eq:2_GravitationalDynamicalTime}
\end{equation}
where $R$ is initial separation between target and projectile centres of mass, $M_{\mathrm{tot}}$ the total mass of the system, and $G$ is the gravitational constant.

The other time constraint we considered was the time required to observe the secondary impact for graze-and-merge collisions. At higher impact parameters as energy increases from perfect merging to hit-and-run, graze-and-merge collisions can occur where the remnants of the projectile have sufficient energy to escape from the initial collision but not enough energy to escape from the gravitational influence of the largest remnant \citep{Genda2012,Emsenhuber2019a}. In such a situation, these projectile remnants will eventually fall back to the target for a secondary collision. Our collisions were run for sufficiently long that we could simulate re-collision orbits of the size of the Hill sphere of the target at $0.4\,\mathrm{au}$ from the central star (see appendix \ref{ap:Graze-and-Merge} for details). We do not include the effects of the central star in our simulations, however, so graze-and-merge collisions are removed from most analyses in the rest of this work.

For all collisions we used the same $6.25\,\mathrm{M_{\oplus}}$ target (the same mass as the middle mass target of our previous work \citealt{Denman2020}). The projectiles were $1,\,2,\,3,\,4\,\&\,5\,\mathrm{M_{\oplus}} $. We chose eight different velocities: $20,25,30,35,40,50,60\,\&\,70 \,\mathrm{km\,s^{-1}}$, the slowest of these is approximately escape velocity (mutual escape velocity was in the range $19.2$--$21.6 \,\mathrm{km\,s^{-1}}$ for all collisions), whereas the fastest is approximately the expected collision velocity of a prograde and retrograde planet at $0.1\,\mathrm{au}$; lower collision velocities are more probable \citep{Gabriel2020} so we carried out a wider range of collisions at low velocities.

 In most previous work on planetary impacts the point at which collision occurs is straightforwardly defined by the time of first contact between the planetary surfaces. Atmospheres, however, do not have a well defined outer ``edge". So that we do not need to define such an edge to the atmosphere, for most collisions we define a collision to occur when the mantle surfaces of target and projectile touch. For highly grazing collisions, in which the projectile only passes through the target atmosphere, however, we define the point of collision to be the point of closest approach. We also use the radius of the surface of the mantle to define the radius of the target, $R_\mathrm{t}$, and projectile $R_\mathrm{p}$. Likewise the impact parameter, $b$, is normalised by the sum of the target and projectile mantle surface radii following
\begin{equation}
    b=\frac{B}{R_{\mathrm{t}}+R_{\mathrm{p}}},
\end{equation}
where $B$ is the vertical distance between the centres of mass of target and projectile as shown in Figure \ref{fig:NM_CollisionGeometry}. This normalisation means that highly grazing collisions where the mantles do not touch will have impact parameter $b>1$, this is in contrast to previous atmosphere-less works on planet collision where $b>1$ meant a collision would not occur. (Impact angle is defined as $\theta_{\mathrm{imp}}=sin^{-1}(b)$, because this is undefined for $b>1$ we do not use it in our analyses.) 

We chose a range of seven different impact parameters roughly evenly spaced, from head-on ($b=0$) to a series of collisions that only passed through the atmosphere at $b=1.3$ (these were $b\,=\,0,\,0.1,\,0.34,\,0.5,\,0.71,\,0.94,\,1.1\,\&\,1.3$). Full data of all of these collisions and their precise collision parameters is given in Tables \ref{tab:CollisionData_1}–\ref{tab:CollisionData_5} in appendix \ref{ap:DataTables}.

To determine the required positions of planets at the beginning of our simulation runs we used the above collision parameters, assumed target and projectile were perfectly spherical and thus well represented by a point mass, and used a Verlet integrator to trace the path of the projectile backwards in time until the centres of mass of both target  and projectile were separated by the distance of five times the sum of their mantle surface radii. This choice of initial separation allowed both objects time to tidally distort before collision. 

For determining the amount of material bound in each remnant after a collision we use the iterative procedure outlined in \citet{Benz1999}. In this method, first the particle lowest in the potential is found, then all the particles gravitationally bound to it are identified. The total mass and centre of mass of this group are calculated and the particles bound to it detected. This process continues until the bound group no longer increases in mass each iteration.

\section{Results}

\begin{figure*}
 \includegraphics[width=2\columnwidth]{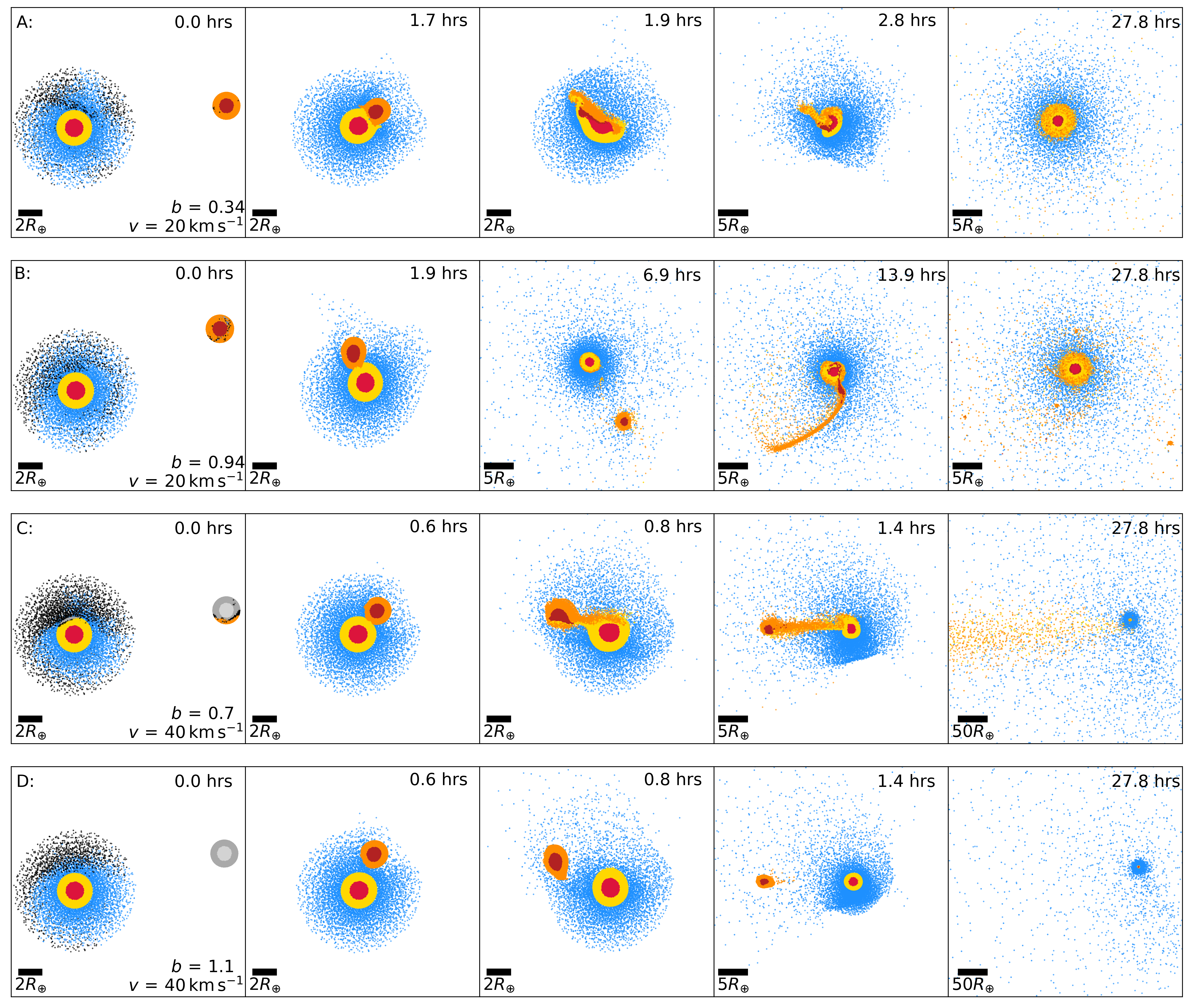}
 \caption{Snapshots showing cross-sectional slices of collisions in progress for four different types of collision between a $6.25\,\mathrm{M_\oplus}$ target and a $2\,\mathrm{M_\oplus}$ projectile. Colours indicate material type, with different shades to distinguish between projectile and target: iron cores are red, forsterite mantle particles are orange/yellow, and atmosphere particles are blue. Colours in the first panel indicate the final location of each particle, black for unbound material, shades of grey are for the second largest remnant, and largest remnant particles retain the colours described previously.  The four different collisions are as follows: A) a head-on like collision 
 -- the majority of target and projectile merge to form a new planet as described in \citet{Denman2020}, the remaining mass is ejected. B) A collision in the transition region between merge and hit-and-run 
 -- 
 this particular example is a graze-and-merge in which the remnants of the projectile undergo a secondary collision with the target; tidal forces tear apart the projectile as it comes back for the second collision resulting in chunks of projectile remaining in orbit. C) An erosive hit-and-run collision -- 
 the mantles of both objects collide causing both to lose mass during the collision; a stream of mantle debris is left between the two objects after the collision.  D) A highly grazing collision --  
 the projectile only passes through the atmosphere of the target, shockwaves eject some of the atmosphere but the projectile and target lose a negligible mass of core and mantle.}
 \label{fig:Res_Snapshots}
\end{figure*}

In this paper we simulate collisions between Super-Earths at a wide array of impact angles, from head-on, to the projectile barely grazing the target's atmosphere, as well as a wide variety of impact speeds, from the escape velocity, to twice the orbital velocity of a target at $0.1\,\mathrm{au}$ (i.e. the expected velocity of a collision between prograde and retrograde planet).
An array of these results is shown in Figure \ref{fig:Res_Snapshots} detailing some of the more common outcomes.

Overall we find that the results are a strong function of specific impact energy, $Q_{\mathrm{R}}$. We see a separation of the results into two regimes. At higher impact parameters and energies a large percentage of the projectile has enough energy to escape the larger target remnant resulting in two large remnant objects with masses of a similar order of magnitude (for example collisions C and D in Figure \ref{fig:Res_Snapshots}). At lower energies and impact angles the projectile remnant cannot escape and we observe behaviour similar to the head-on collisions described in \citet{Denman2020}, where we obtain a single large remnant containing core material from both target and projectile (e.g. collision A in Figure \ref{fig:Res_Snapshots}). We develop scaling laws for these two main regimes, these are summarised in appendix \ref{ap:Prescription}. In between these two regimes we observe a narrow band of graze-and-merge collisions in which the projectile escapes the target on its first collision but does not have enough energy to escape the gravitational influence of the target and thus merges in a later secondary collision (e.g. collision B in Figure \ref{fig:Res_Snapshots}).

\subsection{Largest Remnant Mass}

\begin{figure*}
 \includegraphics[width=\textwidth]{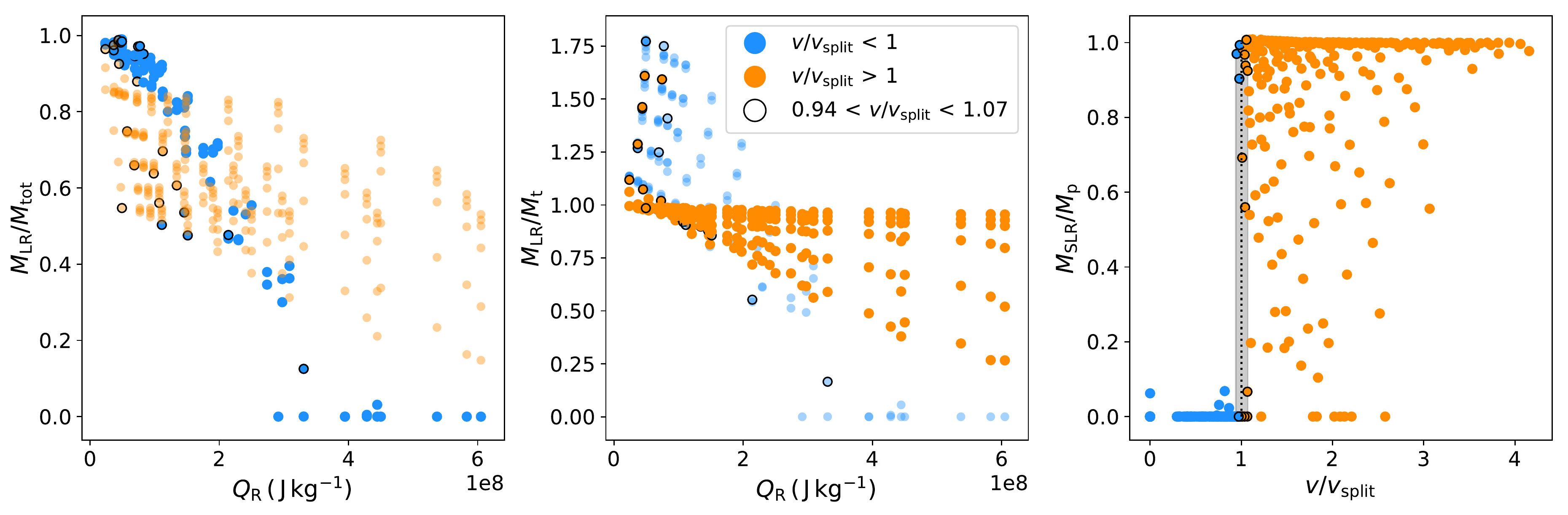}
 \caption{
    \textit{Left:} The mass of the largest remnant as a fraction of total mass compared with impact energy. The collisions with impact velocities below $v_{\mathrm{split}}$, which are well described by \citet{Denman2020}, are coloured in blue, other collisions are orange and partially transparent as their masses are not well described using this normalisation.
    \textit{Centre:} The mass of the largest remnant as a fraction of target mass compared to impact energy. Collisions with impact velocities greater than $v_{\mathrm{split}}$ are coloured orange. The orange points being constrained to a small wedge indicates this normalisation is worth investigating as in it the data for remnant mass caused by collisions of different mass projectiles overlaps (see Figure \ref{fig:Res_EHR}) . Other collisions are coloured blue and made partially transparent.
	\textit{Right:} Second largest remnant mass in terms of the projectile mass compared with the ratio of the impact velocity to the critical velocity, $v_{\mathrm{split}}$. 
	The shaded region shows the values we chose to ignore for our fits to either regime; this selection gives the circled points in all three graphs. 
	     }
 \label{fig:Res_RegimeBoundary}
\end{figure*}

We begin by examining the overall mass of the largest post-collision remnants. The two graphs on the left and centre in Figure \ref{fig:Res_RegimeBoundary} show the dependence of the mass of the largest remnant on the relative specific kinetic energy of the collision, $Q_{\mathrm{R}}$ -- a parameter that has been shown in multiple previous works (e.g. \citealt{Stewart2009,Marcus2009,Leinhardt2012}) to be useful for constructing scaling relations. In the rightmost graph of Figure \ref{fig:Res_RegimeBoundary} we observe two separate collision regimes which result in different masses for the second largest collision remnant. These two collision regimes correspond to situations where there is either one or two resultant large post-collision remnants, i.e. situations in which the cores of target and projectile either merge with one another or survive and continue on separate trajectories after the impact. 
The simulations in the first of these regimes (blue points) result in one single large remnant (leftmost graph, Figure \ref{fig:Res_RegimeBoundary}) formed from the combined cores of both target and projectile and a low or negligible mass secondary remnant. The second of these regimes (orange points) is similar to hit-and-run, with a significant mass secondary remnant present due to the projectile bouncing off the target. The difference between these collisions and the idealised \citet{Leinhardt2012} definition of hit-and-run (i.e. no erosion of the target) is that, due to the presence of the lightly bound atmosphere, some erosion from the target is always observed (centre graph Figure \ref{fig:Res_RegimeBoundary}). The energy cutoff between these two collision regimes is dependent on impact parameter with hit-and-run like collisions at higher $b$ occurring at lower impact energies.

We find that the transition between these two regimes can be well approximated by a simple escape velocity measure,  
\begin{equation}
    v_{\mathrm{split}}=\sqrt{\frac{2GM_{\mathrm{tot}}}{b(R_{\mathrm{p}}+R_{\mathrm{t}})}} .
    \label{eq:v_split}
\end{equation} This equation assumes, firstly, that each planets trajectory is well approximated by describing each planet as a point mass, and also that the impact parameter, $b$, is a good approximation for the distance of closest approach.

These two different collision outcomes are the reason why the remnant mass is best described using two different mass normalisations (left and centre panels in Figure \ref{fig:Res_RegimeBoundary}).
At lower velocities and impact parameters we have a regime with one single large remnant formed by a combination of both projectile and target so we normalise by the combined mass of projectile and target, $M_{\mathrm{tot}}$, as shown by the blue points in the leftmost graph of Figure  \ref{fig:Res_RegimeBoundary}.  The largest remnant mass, $M_{\mathrm{LR}}$, in these collisions behaves like the head-on collisions described in \citet{Denman2020} as shown in Figure \ref{fig:Res_HOfits}. At higher impact energies, when most of the atmosphere is removed, the largest remnant mass in this regime behaves similarly to that described in \citet{Leinhardt2012}, that is decreasing linearly with increasing $Q_{\mathrm{R}}$; at low energies atmosphere removal is less efficient and we, therefore, have a shallower gradient.
The second regime occurs at higher velocities for each impact angle. All largest remnants in the second regime are less massive than the target -- the target always suffers some erosion. We normalise largest remnant mass by target mass in this regime (see orange points in the middle graph of \ref{fig:Res_RegimeBoundary}) because the largest remnant mass can be considered to be less than the target mass by some impact energy dependent erosive factor as will be shown later in Figure \ref{fig:Res_EHR}.

While the rightmost graph in Figure \ref{fig:Res_RegimeBoundary} shows that $v_{\mathrm{split}}$ is a good approximation for the regime boundary, there are a few misdiagnosed results. There are three potential causes for these outliers: firstly the assumption of spherical symmetry, in reality there is some tidal distortion at this point; secondly, graze-and-merge collisions should occur at slightly lower velocity than this boundary, those on long orbits may not have had time to re-collide yet, meaning they may appear as hit-and-run collisions in our data even if their final result would be a merger; or thirdly, material dependent impact effects such as deflection or drag. 

Another confusing factor for the boundary between collision regimes is the possibility of secondary collisions which occur when the secondary remnant does not have the energy to escape the target after collision. These secondary collisions could occur close to the end of the simulation, meaning we would not be able to measure the equilibrium bound mass. For planets orbiting close to the central star the simulation timescale we have used is sufficient for them to escape the Hill sphere of the largest remnant, but because we do not model a central star, they undergo a secondary collision. 
To keep our collision analysis agnostic to orbital distance and to account for the imperfect nature of this regime boundary measure, we have elected to ignore all simulations with velocities in the range $0.94<v_{\mathrm{split}}<1.07$ when fitting.

\subsubsection{Erosive Hit-and-Run}
\label{sec:Res_EHR}
\begin{figure}
 \includegraphics[width=\columnwidth]{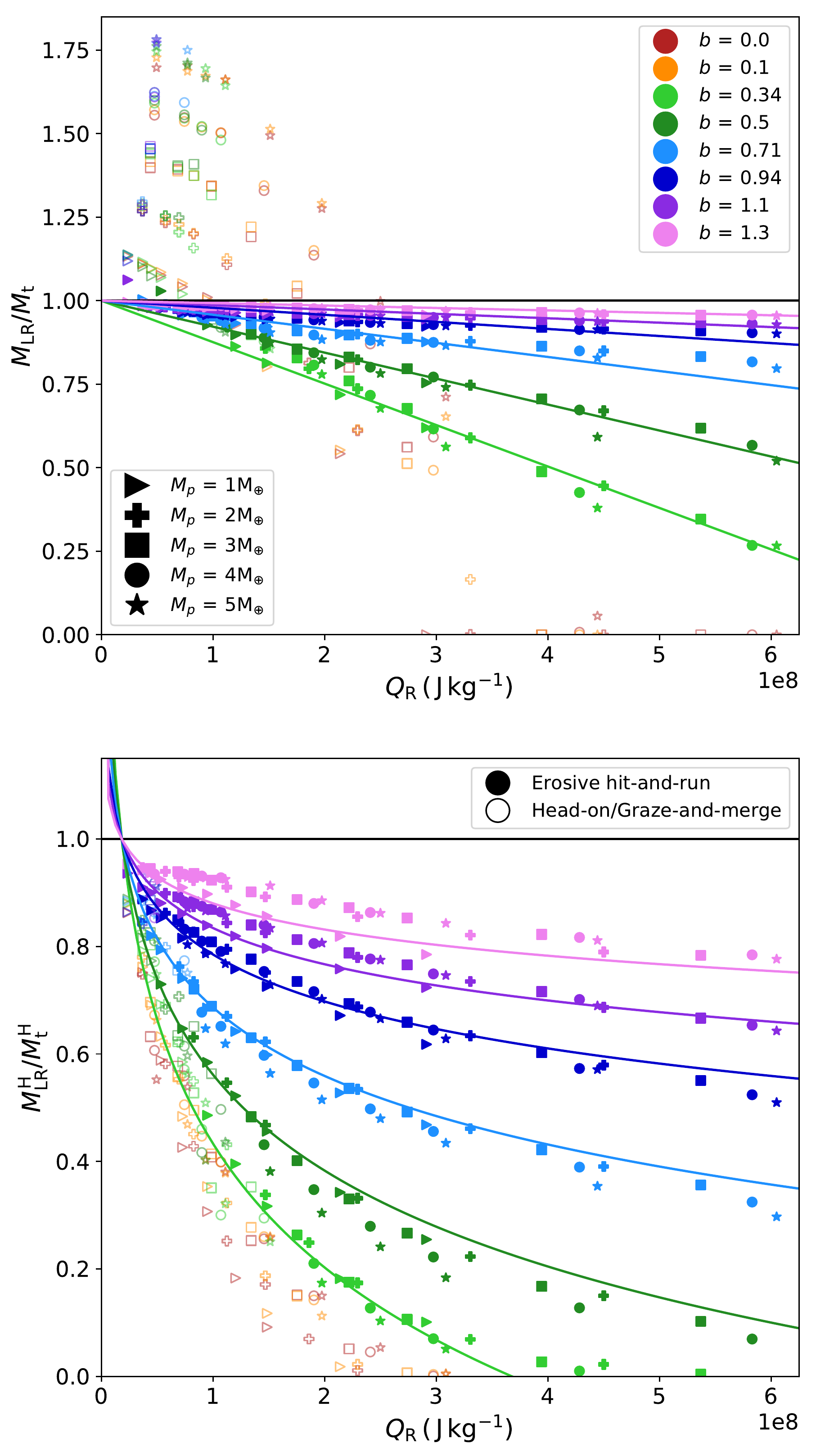}
 \caption{
 \textit{Top:} Mass of the largest remnant relative to the target mass compared with specific impact energy. Erosive hit-and-runs are shown with filled symbols, open symbols indicate head-on or transition region collisions. Projectile mass is indicated by symbol shape; different impact angles are given in different colours. This graph illustrates the projectile mass independence of erosive hit-and-run collisions. The data has been fitted linearly with an intercept of the target mass and a gradient that is a power law function of impact parameter (Equation \ref{eq:MLR_Equation}). The black horizontal line indicates the original target mass.
 \textit{Bottom:} Mass of atmosphere in the largest remnant normalised by the initial atmosphere mass of the target compared to $Q_{\mathrm{R}}$. The fit shown by the solid lines is a power law fit with a shared  minimum energy for atmosphere loss $Q_{\mathrm{0}}$, and a power law coefficient that is itself a power law function of impact angle (Equation \ref{eq:Res_LRAtmosphereFit}).
 }
\label{fig:Res_EHR}
\end{figure}

For collisions with a velocity above $v_{\mathrm{split}}$ the projectile rebounds off the target causing erosion in one or both; following \citet{Gabriel2020} we call such collisions erosive hit-and-runs.

In these erosive hit-and-run collisions we observe an approximately linear decrease in largest remnant mass (normalised by the target mass) with increasing impact energy (top graph, Figure \ref{fig:Res_EHR}). This linear relationship is independent of projectile mass, but with a gradient that decreases with increasing impact parameter, meaning that more grazing collisions remove less material.

The gradient of this linear relationship $\Gamma_{\mathrm{EHR}}$ can be shown to have a power law dependence on impact parameter (see appendix \ref{ap:EHR_Construction}).
If we fit to all of the impact parameters simultaneously we obtain:
\begin{equation}
    \frac{M_{\mathrm{LR}}}{M_{\mathrm{t}}}=-10^{\left(-1.28\pm0.02\right)b-8.47\pm0.01}Q_{\mathrm{R}}+1 .
    \label{eq:MLR_Equation}
\end{equation}
We have set this model to pass through $M_{\mathrm{LR}}=M_{\mathrm{t}}$ at $Q_{\mathrm{R}}=0$ so that we have no mass loss for zero input energy. A fit with a loose intercept (see appendix \ref{ap:EHR_Construction}) gives an intercept slightly below one indicating that low energy mass loss is potentially non-linear, despite this we have still used a linear fit so as to provide the simplest description of the results.

Decrease in mass ejected with increasing impact parameter is observed in many previous collision works on atmosphere-less collisions \citep[e.g.][]{Leinhardt2012,Movshovitz2016}. To determine the effects of the atmosphere on the impact angle dependence of the largest remnant we compare our results to those of \citet{Leinhardt2012}. They found that their observed largest remnant mass could be well described by 
a simple relation based on the geometry of the impact that described the dependence of the amount of impact energy the projectile deposits in the target on the impact parameter. 
They show the effective specific kinetic energy of a collision at a particular impact angle $Q^{'}_{\mathrm{R}}$, is related to the specific kinetic energy of a head-on collision at the same mass and velocity, $Q_{\mathrm{R}}$, by
\begin{equation}
Q^{'}_{\mathrm{R}}=\frac{\mu_{\mathrm{\alpha}}}{\mu}Q_{\mathrm{R}},
\end{equation}  
where $\mu$ is the reduced mass and $\mu_{\mathrm{\alpha}}$ a modified version of the reduced mass determined from the mass of the projectile that interacts with the target. $\mu_{\mathrm{\alpha}}$ is given by 
\begin{equation}
\mu_{\mathrm{\alpha}}=\frac{ \alpha M_{\mathrm{p}} M_{\mathrm{t}} }{ \alpha M_{\mathrm{p}} + M_{\mathrm{t}} },
\end{equation}
where $\alpha$ is found by calculating the fraction of the projectile that is below the uppermost point of the target at the point of impact (see figure 2 in \citealt{Leinhardt2012}):
\begin{equation}
\alpha=\frac{3R_{\mathrm{p}}l^{2}-l^3 }{4R_{\mathrm{p}}},
\end{equation}
 $l$ here is  the vertical distance between the base of the projectile and the top of the target. Note that $l=(R_{\mathrm{p}}+R_\mathrm{t})(1-b)$. In this way the energy involved in the oblique impact is always smaller than (or equal to) the energy of the head-on impact.

This correction to the impact energy can be related to our gradient, $\Gamma_{\mathrm{EHR}}=-10^{-1.28b-8.47}$, above by considering the simultaneous equations of equation \ref{eq:MLR_Equation} and the equivalent equation we would get with the corrected impact energy, $Q^*_{\mathrm{R}}$:
\begin{equation}
\frac{M_{\mathrm{LR}}}{M_{\mathrm{t}}}=-\Gamma_{\mathrm{0}} Q^*_{\mathrm{R}}+1.
\end{equation}
This leaves us with
\begin{equation}
Q^*_{\mathrm{R}}=\frac{\Gamma_{\mathrm{EHR}}(b) }{\Gamma_{\mathrm{0}}}Q_{\mathrm{R}},
\end{equation}
or in other words the reduced mass ratio we expect to observe in our collisions should be directly proportional to $\Gamma_{\mathrm{EHR}}$ and some constant scaling factor $\Gamma_{\mathrm{0}}$,
\begin{equation}
\frac{\Gamma_{\mathrm{EHR}}(b)}{\Gamma_{\mathrm{0}}}=\frac{\mu_{\mathrm{\alpha}}}{\mu}.
\end{equation}

\begin{figure}
 \includegraphics[width=\columnwidth]{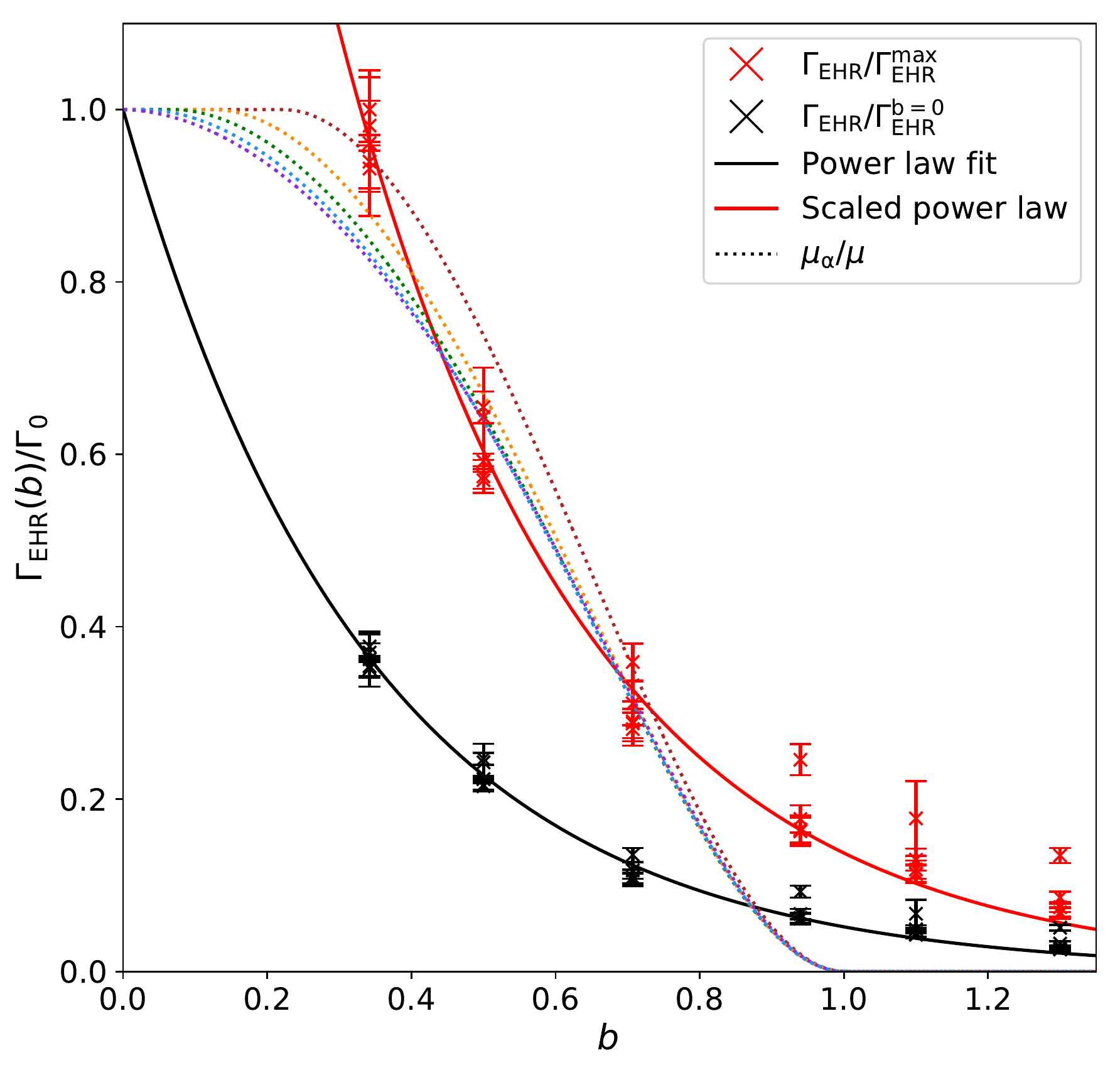}
 \caption{
 Two different scalings of the total mass loss gradient as a function of impact parameter. 
 The scaling in black is with respect to the predicted gradient at $b=0$ from the fit in equation \ref{eq:MLR_Equation}. The data points are from the measured gradients for linear fits to individual projectile masses and impact parameters (see appendix \ref{ap:EHR_Construction}). The red line and data points are scaled with respect to the greatest measured gradient.
 These are compared to our modified version of \citet{Leinhardt2012}'s reduced mass ratio modification to $Q_{\mathrm{R}}$. Our results at the lower impact angles follow a pattern consistent with \citet{Leinhardt2012}, but they diverge at higher impact angles due to the presence of an atmosphere.}
 \label{fig:Res_mintfits}
\end{figure}

The exact correction from \citet{Leinhardt2012} cannot be used for our collisions with an atmosphere because atmospheres do not have a defined outer edge, instead they decrease roughly exponentially in density with distance from their base. Instead we assume the radius of the target is the outer radius of its mantle. The relation between reduced mass ratio and impact parameter predicted using this modified model is shown by the coloured lines in Figure \ref{fig:Res_mintfits}.

When $\Gamma_{\mathrm{EHR}}(b)/\Gamma_{\mathrm{0}}$ is scaled up such that $\Gamma_{\mathrm{0}}$ is the largest measured gradient  ($\Gamma_{\mathrm{EHR}}(b=0.34)$, red points in Figure \ref{fig:Res_mintfits}), the low impact parameter collisions ($b\leq0.8$) agree well with the modified version of the \citet{Leinhardt2012} interacting mass measure using the mantle surface radius as the target radius (see Figure \ref{fig:Res_mintfits}). This agreement indicates that the corrected \citeauthor{Leinhardt2012} prediction of the relation between impact parameter and impact energy of the interacting mass still works well for collisions in which the majority of interaction between projectile and target is between core and mantle, even though the prediction does not include the effects of the atmosphere. 
The shock caused by the projectile passing through the atmosphere however means that the amount of material removed at higher impact angles is greater. 
The atmosphere-less bodies in \citet{Leinhardt2012} would no longer impact one another at impact parameters greater than one so there would be no energy deposition and no mass loss, whereas, due to our definition of the point of impact, the projectile dives into the target's atmosphere at these impact parameters (1.0 -- 1.3) and we observe atmospheric erosion. 
This atmosphere material is removed preferentially to the mantle for two reasons: firstly, it is significantly less tightly bound than the core and mantle material, secondly, the impedance mismatch between atmosphere and mantle mean that any shockwaves in the mantle caused by atmospheric shocks are substantially weaker than the initial shock that caused them. This low level of mantle interaction is especially true for highly grazing ($b>1$) collisions.

\subsubsection{Atmosphere Loss in the Erosive Hit-and-Run Regime}
\label{sec:Res_atmosEHR}

To understand how much atmosphere gets removed in these erosive hit-and-run collisions between atmosphere rich and atmosphere-less Super-Earths we compare the specific impact energy, $Q_{\mathrm{R}}$, with the fraction of atmosphere remaining on the largest remnant after a collision, $M_{\mathrm{LR}}^{\mathrm{H}}/M_{\mathrm{t}}^{\mathrm{H}}$, where $M^{\mathrm{H}}$ is the mass of the hydrogen atmosphere and the subscripts $\mathrm{t}$ and $\mathrm{LR}$ indicate the target and the largest remnant respectively. Figure \ref{fig:Res_EHR} shows that the amount of atmosphere in the largest remnant decreases with increasing impact energy and also with decreasing impact parameter. The amount of extra material removed for a small increase in energy also decreases with increasing energy. The data for each separate impact parameter thus follow convex curves downwards.

In the erosive hit-and-run regime each of the separate projectile masses (shown by different shapes in Figure \ref{fig:Res_EHR}) all follow the same loss curve for each impact parameter. This means that any projectile mass dependency in the atmosphere mass loss from these impacts is characterised by $Q_{\mathrm{R}}$.

In all collisions the amount of material ejected is proportional to the difference between the input kinetic energy of the impact, and the gravitational potential energy of the material being ejected, minus any losses due to heat generating processes. Physically the lack of any additional dependence on projectile mass beyond specific impact energy is likely due to the projectile rebounding off the target sufficiently quickly that the material it causes to be ejected from the target only needs to escape the target's gravitational potential. It should be noted that for projectile-target mass ratios smaller than the range tested here ($0.16-0.8$) one might expect this symmetry to be broken, as smaller projectiles have lower momenta and are thus likely to deposit more of their kinetic energy in the atmosphere. This phenomena is likely the reason why the $1\,\mathrm{M_{\oplus}}$ results (the triangles in Figure \ref{fig:Res_EHR}) are not in agreement with the higher masses for our higher impact parameter simulations.

Initial power law fits for each impact parameter (appendix \ref{ap:EHR_Construction}) crossed the initial atmosphere mass at a non-zero energy, this suggests that there is a minimum input energy required for atmosphere to be ejected from the planet. The precise value for this minimum energy, $Q_{\mathrm{0}}$, for each of the initial fits is given in graph C Figure \ref{fig:ap_EHRconst} and is due to the energy required to accelerate atmosphere particles to a velocity where they can escape the largest remnant's Hill sphere. As such the precise value is likely to be dependent both on how efficiently collision energy is spread between atmosphere particles and also potentially for simulations, their resolution.

The power law coefficient was also found to itself have a power law dependence on impact parameter (see appendix \ref{ap:EHR_Construction}). Fitting for all projectile masses and impact angles simultaneously with the function
\begin{equation}
    \frac{ M_{\mathrm{LR}}^{\mathrm{H}} }{ M_{\mathrm{t}}^{\mathrm{H}}  }=-10^{\delta b+\epsilon} \log_{10}\left(\frac{Q_{\mathrm{R}}}{Q_{\mathrm{0}}}\right)+1,
  \label{eq:Res_LRAtmosphereFit}
\end{equation}
(solid lines in the bottom graph of Figure \ref{fig:Res_EHR}) we obtained values of $\delta=-0.70\pm0.01$, $\epsilon=0.13\pm0.01$ and $Q_{\mathrm{0}}=(1.8\pm0.1)\times10^6\,\mathrm{J\,kg^{-1} }$.

The physical meaning of the impact parameter dependent power law coefficient, $\gamma_{\mathrm{atmos}}=-10^{\delta b+\epsilon}$, is related to the efficiency of the collision in removing atmosphere from the target. Collisions closer to head-on remove more atmosphere than more highly grazing ones as a greater amount of the projectile's momentum can be transferred to the target's atmosphere. In addition, for collisions with low $b$ the projectile causes shockwaves that travel through the core and mantle, which can cause atmosphere ejection from all over the planet not just the parts close to the trajectory of the projectile. Atmospheric shockwaves from high $b$ impacts, on the other hand, are unlikely to cause strong shocks in the mantle due to the impedance mismatch between atmosphere and mantle.

\subsubsection{The Head-On-Like Regime}
\label{sec:3_HeadOn}

\begin{figure}
 \includegraphics[width=\columnwidth]{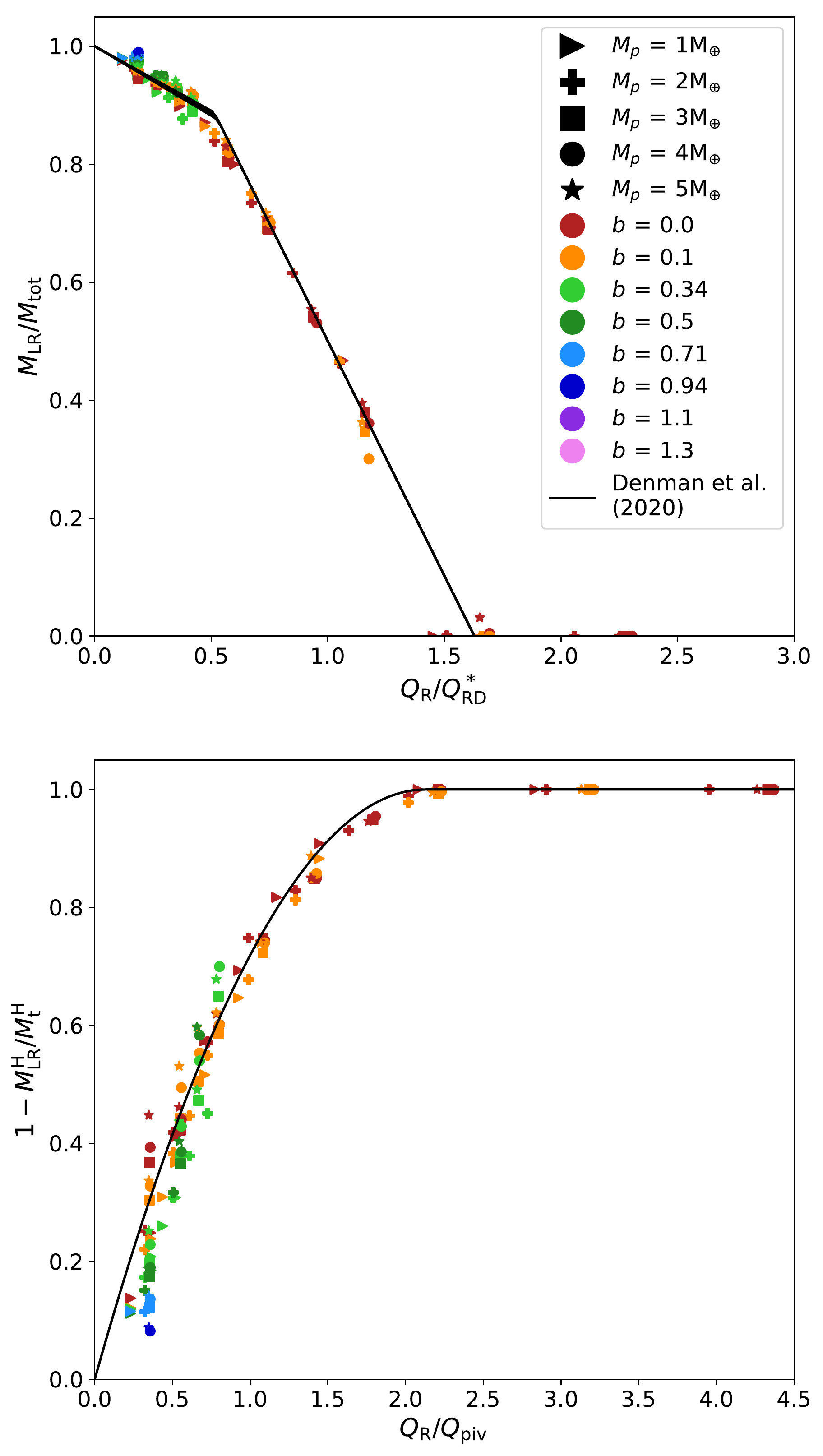}
 \caption{
    \textit{Top:} Largest remnant mass as a function of normalised specific impact energy for all collisions in the head-on-like regime. Results are compared to the model from \citet{Denman2020} (the black line). We observe both the transition energy $Q_{\mathrm{piv}} $ between atmosphere dominated loss at low energies and core dominated loss at high energies, and also the loss gradients to be in close agreement with this prediction. 
	\textit{Bottom:} The fraction of atmosphere lost by the largest remnant for all collisions with velocity below $v_{\mathrm{split}}$ (head-on-like) compared with impact energy normalised by $Q_{\mathrm{piv}}$ from \citet{Denman2020} (see appendix \ref{ap:Prescription} for calculation). The black line shows their predicted atmosphere loss fraction. The two results agree well with one another. 
	}
 \label{fig:Res_HOfits}
\end{figure}

For collisions below the critical velocity, $v_{\mathrm{split}}$, we compare the mass in the largest remnant to the specific impact energy normalised by the catastrophic disruption threshold $Q_{\mathrm{RD}}^*$ outlined in \citet{Denman2020}, as shown in Figure \ref{fig:Res_HOfits}. We observe an initial shallow decrease in mass with increasing impact energy, which sharply increases at about $0.5 Q_{\mathrm{RD}}^*$; this is strongly consistent with the model derived in \citeauthor{Denman2020} (black lines in Figure \ref{fig:Res_HOfits}), as such we label all these collisions as `head-on-like'. \citet{Denman2020} identify this transition with the minimum energy at which mantle will be excavated by the impact in addition to atmosphere.

We have also compared the fraction of atmosphere lost from the largest remnant for this set of head-on-like collisions to our predictions in \citet{Denman2020} (see bottom panel, Figure \ref{fig:Res_HOfits}). Again the results are strongly consistent with the predicted values. There is, however, an impact angle dependent divergence from predicted atmosphere loss which increases at low impact energies. More grazing collisions ($b\gtrsim 0.3$) remove less atmosphere than predicted by the model (blue and green points). Glancing collisions between planets without atmospheres (such as those detailed in \citet{Leinhardt2012} section 3.1.2) have been shown to have less efficient energy deposition at higher impact parameters. A less energetic shockwave in the core means less atmosphere being pushed away by the shock as it reaches the target surface again on the other side of the planet. 

This consistency between the results of this work and the model from \citet{Denman2020} occurs despite this paper using a significantly increased atmosphere pseudo-entropy, $1.3\times10^{12}\mathrm{Ba\,g^{-\gamma}\,cm^{3\gamma}}$ as opposed to $5\times 10^{11}\mathrm{Ba\,g^{-\gamma}\,cm^{3\gamma}}$, which resulted in a warmer, more extended atmosphere. Looking at these collisions in terms of energy, this consistency is likely because although the higher pseudo-entropy means the outermost atmosphere particles are higher up the potential well, the difference in energy required for them to be ejected is small compared to the differences in how efficiently the impact energy is spread between particles.

\subsection{Remnant Composition}

\begin{figure}
 \includegraphics[width=0.85\columnwidth]{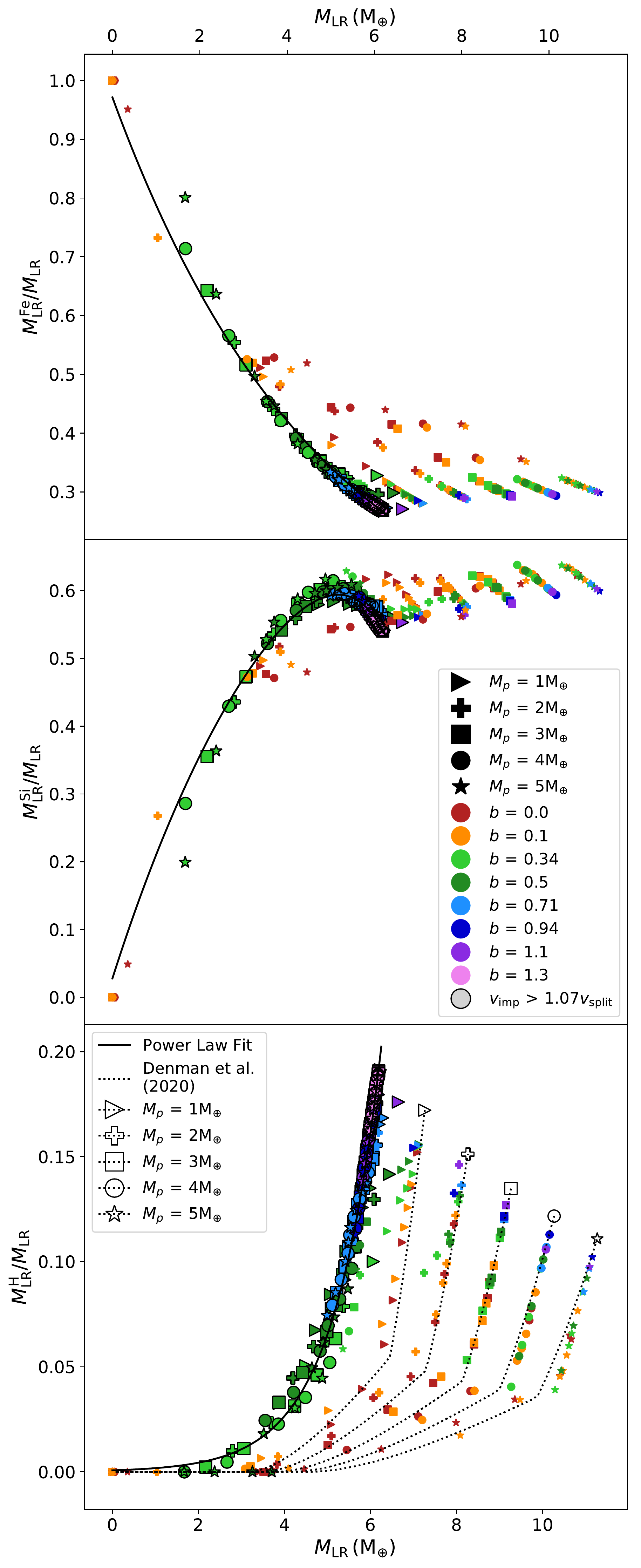}
 \caption{
    \textit{Top:} Mass fraction of iron in the largest remnant as a function of the remnant mass. Iron fraction increases at low remnant masses because collisions preferentially remove material that is less well bound.
    \textit{Middle:} Silicate mass fraction of the largest remnant. The mantle is less tightly bound than core material yet more strongly bound than atmosphere, we thus observe a turnover at the point where the majority of atmosphere has been eroded and mantle erosion begins in earnest.
	\textit{Bottom:} The fraction of the largest remnant that is hydrogen atmosphere. Atmosphere is the least strongly bound part of the planet and thus is removed preferentially. We also show the predictions for head-on collisions from \citet{Denman2020} as dotted lines (open symbols at the end represent the projectile mass of the prediction). Colours and symbols have the same meanings as in previous Figures. For each of these graphs we have fit the $v_{\mathrm{imp}}>v_{\mathrm{split}}$ data with a power law (solid black lines).
	     }
 \label{fig:Res_MaterialFractions}
\end{figure}

Preferential erosion or accretion of planet components during collisions will result in compositional change. 
Figure \ref{fig:Res_MaterialFractions} shows the fraction of largest remnant that is comprised of each constituent material after a collision. The original mass fractions were, $26\%$ iron, $53\%$ forsterite and $20\%$ hydrogen for the target, and, $33\%$ iron and $67\%$ forsterite for the projectile.
We observe two different patterns: head-on-like collisions with $v<v_{\mathrm{split}}$ tend to clump around separate curves dependent on projectile mass; whereas the erosive hit-and-run collisions, for which $v>v_{\mathrm{split}}$, tend to all follow one single curve, this is true for the fractions of all material types (core, mantle, and atmosphere). The single line for erosive hit-and-runs is due to the collisions only eroding material from the target and barely depositing any.

Comparing the three graphs in Figure \ref{fig:Res_MaterialFractions}, looking at the erosive hit-and-run collisions, we see iron core content increasing with decreasing largest remnant mass (top), Hydrogen atmosphere content by comparison decreases with decreasing final mass (bottom), while forsterite mantle content increases with mass removed initially it reaches a turning point at $5.2\pm0.2\,\mathrm{M_\oplus}$ and starts to decrease again as more material is removed (middle). Whether a material fraction increases or decreases with decreasing final remnant mass depends upon if material is preferentially removed in an impact or not. In our simulations, first the lightly gravitationally bound outer atmosphere is removed, then the more tightly bound mantle layer underneath, and finally the strongly bound iron core material. This explains the turning point in mantle fraction, at lower final masses the majority of atmosphere has already been removed and mantle is then removed preferentially to iron core, but at higher masses atmosphere is still being removed preferentially to mantle. 

An important result to take from all erosive hit-and-runs following the same pattern of final material fraction, is that the atmosphere fraction after an erosive  hit-and-run collision can be calculated from the final remnant mass and vice versa.

The material fraction results were fitted with power laws; for the iron core we obtained:
\begin{equation}
    f_{\mathrm{Fe}}=10^{(-0.091\pm0.002)M_{\mathrm{LR}}-0.01\pm0.01},
    \label{eq:Res_fFe}
\end{equation}
while for the hydrogen atmosphere we obtained,
\begin{equation}
    f_{\mathrm{H}}=10^{(0.39\pm0.01)M_{\mathrm{LR}}-3.14\pm0.05},
    \label{eq:Res_fH}
\end{equation}
finally we combine equations \ref{eq:Res_fFe} and \ref{eq:Res_fH} for the fraction of the largest remnant that is silicate mantle:
\begin{equation}
    f_{\mathrm{Si}}=1-f_{\mathrm{FE}}-f_{\mathrm{H}}.
\end{equation}
These relations are shown in Figure \ref{fig:Res_MaterialFractions} with solid black lines.

This result implies that, in the erosive hit-and-run regime, no matter the projectile-target mass ratio (within the range tested) a collision that results in the same largest remnant mass (i.e. the same normalised impact energy) will result in the same material fractions.

For the atmosphere fraction we have also added lines detailing the prescription from \citet{Denman2020} --  the dotted lines with open symbols representing their projectile mass in the bottom panel of Figure \ref{fig:Res_MaterialFractions}. 
Because our head-on results show strong consistency with the model for total and atmospheric mass in the largest remnant, we again observe consistency here, with some deviation occurring for the $b=0.34$ collisions which are close to the transition region,  $v\approx v_{\mathrm{split}}$.

\subsection{The Second Largest Remnant}

\begin{figure}
 \includegraphics[width=\columnwidth]{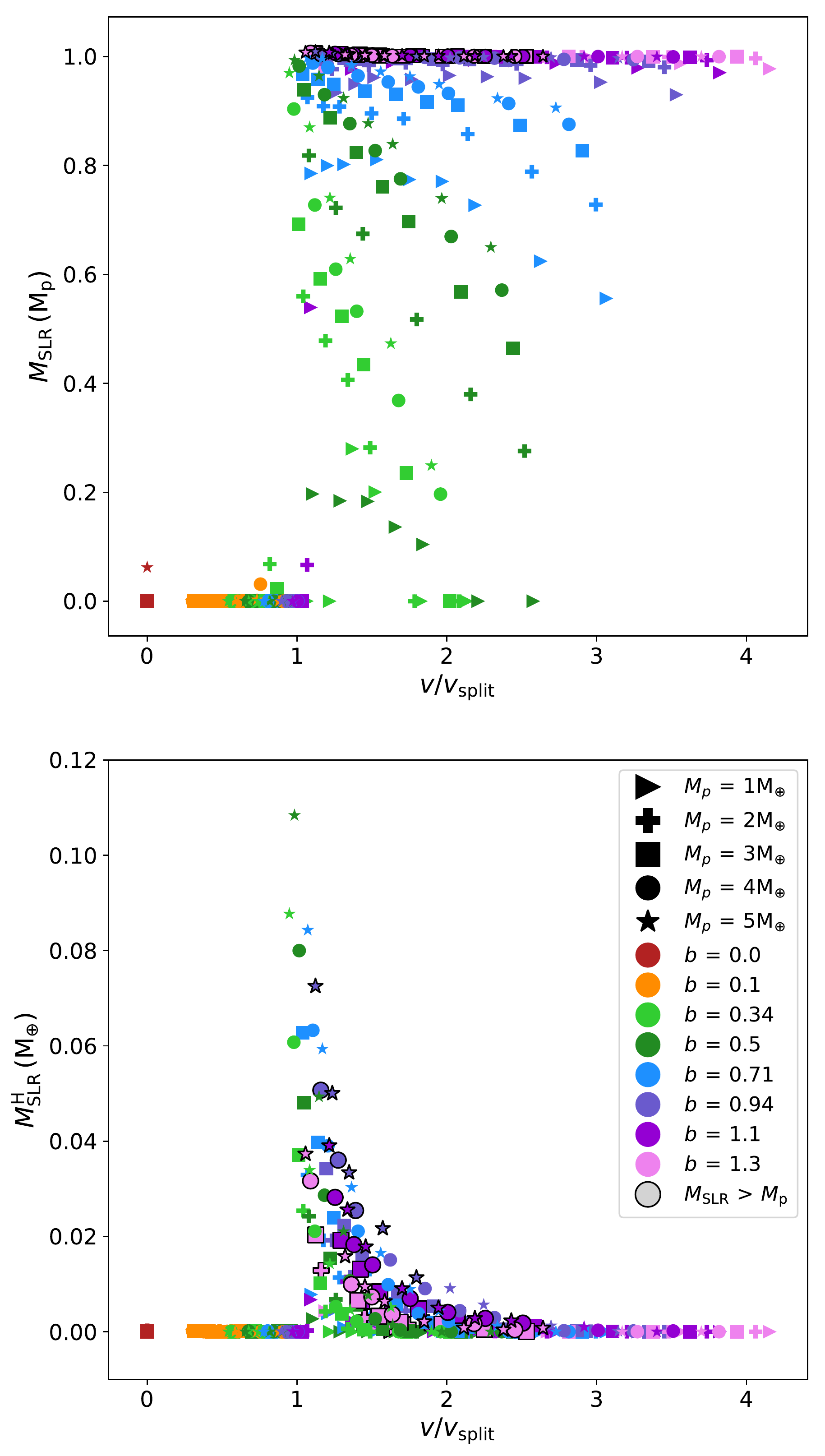}
 \caption{
    \textit{Top:} Mass of the second largest remnant as a function of impact velocity in terms of the transition velocity, $v_{\mathrm{split}}$, showing the dependence on both projectile mass (different symbols) and impact parameter (different colours). More grazing impacts (e.g. blue and purple) result in less material lost from the projectile, as do lower velocity impacts, although sufficiently low velocity impacts will result in a merger ($v<v_{\mathrm{split}}$). Points with a black outline indicate those collisions in which the second largest remnant has increased in mass, typically by accreting atmosphere from the target.
	\textit{Bottom:} The mass of target atmosphere accreted on to the second largest remnant as a function of impact velocity. The amount of accreted atmosphere peaks at a low energy close to the edge of the grazing erosion regime at $v=v_{\mathrm{split}}$.
	     }
 \label{fig:Res_SLR_vs_Qr}
\end{figure}

The top panel of Figure \ref{fig:Res_SLR_vs_Qr} shows the dependence of the mass of the second largest remnant on impact velocity. At low relative velocities and impact angles we observe head-on like behaviour with no secondary remnant of more than a few particles in most cases. The most massive second largest remnants we observe are found at velocities just above the transition velocity, $v_{\mathrm{split}}$.
As the energy increases the projectile gets more and more eroded by the impact. For lower impact parameters, $b \sim 0.3-0.5$, eventually at sufficiently high energy the projectile will get disrupted. The degree to which the projectile is eroded is dependent on impact parameter, with more grazing impacts requiring more energy to remove the same amount of material. For the highly grazing impacts ($b>1$) the amount of mass lost is negligible, and at lower impact energies the projectile can even gain mass. This mass gain comes from accretion of some of the comparatively lightly bound atmosphere of the target as the projectile passes through it (these collisions are highlighted by black outlines in Figure \ref{fig:Res_SLR_vs_Qr}).

The bottom panel of Figure \ref{fig:Res_SLR_vs_Qr} shows the amount of atmosphere that is accreted onto the secondary remnant from the target. Like for the total secondary remnant mass, accreted atmosphere mass peaks at low impact velocities just above the critical impact velocity ($v_{\mathrm{split}}$). The amount of atmosphere accreted decreases a lot more sharply after this point than the total mass of the secondary remnant though, with little to no atmosphere being accreted at velocities twice that of peak accretion. 

All second largest remnants which have greater mass than the projectile have accreted atmosphere from the target during the collision. The outermost target atmosphere particles are loosely bound so the gravitational pull of the projectile can be sufficient to pull them away if they are not completely ejected by atmospheric shocks. This effect is likely exaggerated by the projectiles in our study being atmosphere-less, this means target atmosphere particles can approach closer to the projectile and thus experience a stronger gravitational pull towards it.

The collisions where mass increases also occur at the highest impact parameters where the projectile only barely grazes the target mantle, if at all, as collisions where the mantles collide are sufficiently more erosive to the projectile that it will erode more than it will accrete. The collisions in which the secondary remnants accrete the most atmosphere, however, tend to be those of lower impact parameters (around $0.3<b<0.75$) where more of the core and mantle of the projectile is eroded, but the projectile passes through more of the atmosphere and thus can accrete more.

\section{Discussion}

\subsection{The Erosive Hit-and-Run Regime}

An important difference between collisions involving planets with and without an atmosphere is that, due to the lightly bound nature of the upper atmosphere, there is a very low probability of getting a `true' hit-and-run collision in which target and projectile mass are both affected negligibly. Instead we get `erosive hit-and-runs' \citep{Gabriel2020} where collisions resulting in two large remnants also erode the target.

In this erosive hit-and-run regime we normalise the mass of the largest post-collision remnant with respect to target mass as opposed to total mass. 
This is because in erosive hit-and-run collisions the largest remnant is formed from the partially eroded target with the remnants of the projectile rebounding, as opposed to projectile and target cores combining similar to what happens in a head-on collision.

Our results for the mass of the largest remnant in the erosive hit and run regime show negligible dependence on the projectile mass for our particular target mass. Because of this lack of projectile mass dependence we have elected not to describe our results in terms of accretion efficiency \citep{Asphaug2010a}. 
In this regime the dependence of the amount of material excavated by the projectile on projectile mass is completely characterised by the specific impact energy, $Q_\mathrm{R}$. The efficiency of this excavation is a strong function of impact parameter, with grazing impacts removing a lower proportion of the target's mass. Resolution test simulations showed minimal differences to the standard resolution this analysis is based on, being on average within $2\%$ (see Appendix \ref{ap:ResolutionTesting}).  
This relation is also seen in  the data provided by \citet{Gabriel2020} for atmosphere-less silicate-iron planets.

\subsection{The Transition to Hit-and-Run}

\begin{figure}
 \includegraphics[width=0.94\columnwidth]{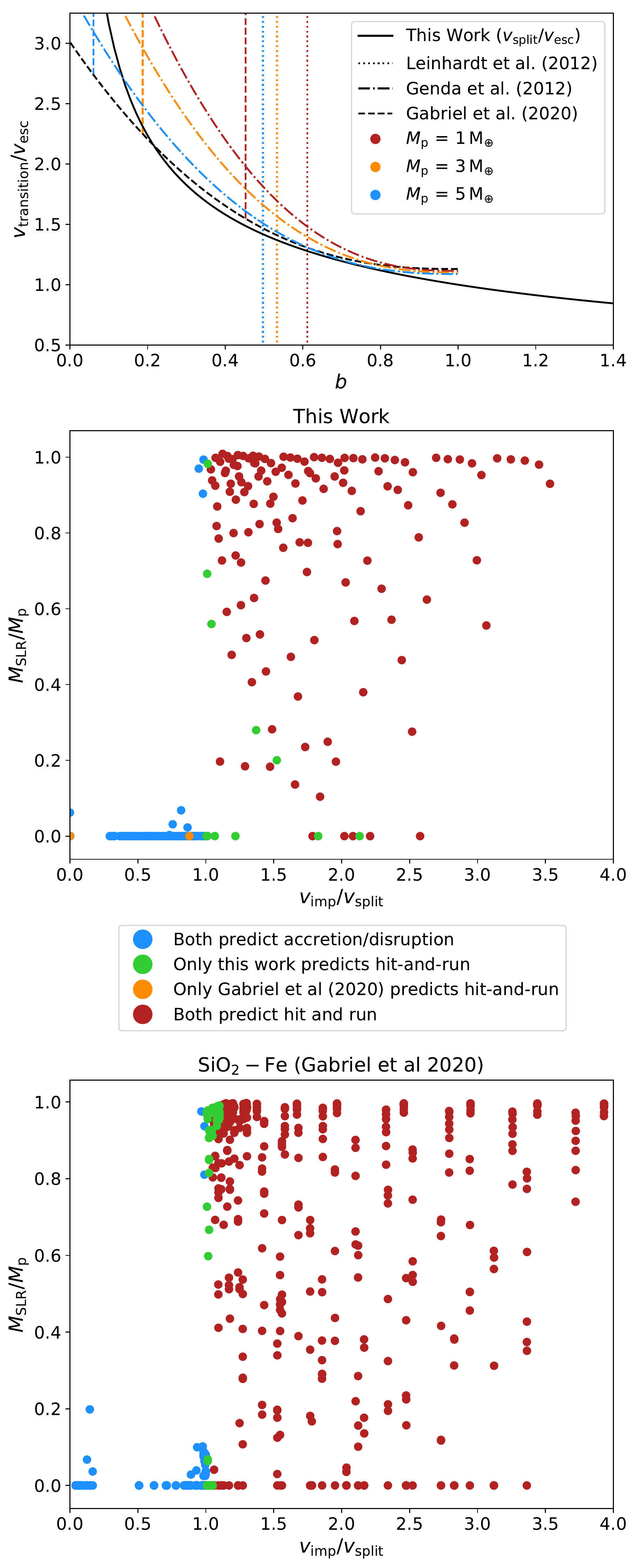}
 \caption{ 
    \textit{Top:}  
    Different methods of calculating the velocity of transition between merging and 'hit-and-run' and the dependence on $b$. Our model provides a close match to the \citet{Gabriel2020} prediction for impact angles greater than their predicted minimum hit-and-run impact angle (dashed lines).
    \textit{Middle:} Comparison between the \citeauthor{Gabriel2020} method for determining the erosive hit-and-run transition and our model applied to our simulations. Our results are in agreement apart from some collisions we classify as erosive hit-and-run (green points). 
	\textit{Bottom:} A comparison between our method for determining the erosive hit-and-run transition and the \citeauthor{Gabriel2020} method, applied to their two component iron-silicate planets. Again our predictions identify more erosive hit-and-run collisions.
	}
 \label{fig:disc_TransitionRegionLocation}
\end{figure}

We find the transition between the merging and hit-and-run collision regimes to be strongly related to the mutual escape velocity of the two bodies at the point of closest approach, $v_{\mathrm{split}}$ defined in equation \ref{eq:v_split}.

Earlier works on impacts such as \citet{Leinhardt2012} use the \citet{Asphaug2010a}  grazing angle measure of $b_{\mathrm{crit}}=R_{\mathrm{t}}/\left(R_{\mathrm{p}}+R_{\mathrm{t}}\right)$ to approximate the transition to hit-and-run. $b_{\mathrm{crit}}$ is the boundary impact parameter where the velocity vector of the projectile's centre of mass no longer intersects the target.
\citet{Gabriel2020} show that for larger planets $\approx1\,\mathrm{M_\oplus}$, where there can be significant stratification of density due to both gravity and material types, this can cause the boundary between merging and hit-and-run to occur at substantially lower impact parameters. The smallest planets in this paper are $1\,\mathrm{M_\oplus}$ and the targets all have large hydrogen atmospheres which are significantly less dense than the cores, we thus observe a transition to hit-and-run collisions at less than $b_\mathrm{crit}$.

We use a different parametrisation of this boundary, one that has a simple functional form and physical motivation, i.e. the velocity required for the secondary remnant to escape the system from its point of closest approach ($v_{\mathrm{split}}$). The top panel in Figure \ref{fig:disc_TransitionRegionLocation} shows comparisons between our model and the models from \citet{Gabriel2020}, \citet{Genda2012} and \citet{Leinhardt2012} where we have replaced the target full radius with that of its surface due to the presence of an atmosphere. In the simplest model, like that used by \citet{Leinhardt2012}, there is a sharp (mass ratio dependent) impact parameter cutoff above which all collisions are hit-and-run (dotted vertical lines). Both our model and that of \citet{Genda2012} (solid lines and dash-dotted lines respectively) describe the transition to the hit-and-run regime as impact parameter dependent, occurring at increasingly higher velocities at lower impact parameters, following a convex function. The model of \citet{Gabriel2020} (dashed lines) uses aspects of both descriptions, below a particular impact parameter threshold hit-and runs will not occur, but above that they will occur if they are above some transition velocity which decreases as a convex function of increasing impact parameter.

The bottom two panels compare which collisions both our method and that of \citet{Gabriel2020} classify as merge or hit-and-run for both our work and the two component bodies from their work. The central condensation parameter $\Lambda$ for each projectile-target pair used our simulations was calculated to be between $0.35$ for the smallest projectile and $0.58$ for the largest. Red points are collisions that both methods classify as erosive hit-and-run, blue points are collisions in which both methods predict accretion or disruption, and green points are collisions that only this work classifies as erosive hit-and-run. Only one point (given in orange) was classified as erosive hit-and-run by \citeauthor{Gabriel2020} but not by our model. The two panels show the two methods give broadly consistent results, with $v_{\mathrm{split}}$ from this work classifying more collisions as hit-and-run.

The minimum possible velocity of a collision between two planets with $b<1$ is their mutual escape velocity, if the two planets are in isolation this is $v_{\mathrm{esc}}\,=\,\sqrt{2GM_{\mathrm{tot}}/(R_{\mathrm{p}}+R_{\mathrm{t}})}$. For highly grazing collisions ($b>1$) the escape velocity from the point of closest approach is $v_{\mathrm{split}}$ which implies that they can only be erosive hit-and-run collisions.
In our simulations, however, we observe some highly grazing collisions to result in mergers. These highly grazing mergers occur because the atmosphere also has a drag force on the projectile as it passes through it, if the projectile is moving slowly enough, this drag force can be sufficient to cause it to become a graze-and-merge collision. In the real world collisions slightly slower than mutual escape velocity  can also occur due to the presence of the other bodies in the system, if we also consider the presence of the central star for example we only need to consider the projectile starting at zero velocity at the edge of the Hill sphere (rather than at infinity).

Material properties may affect the accuracy of $v_{\mathrm{split}}$ as a predictor of the transition between head-on-like and erosive-hit-and-run collisions, for example by changing how much drag or deflection the projectile undergoes as it passes through the target.
$v_{\mathrm{split}}$ seems to work well for both simulations despite differences in equation of state and the presence of an atmosphere. This agreement is likely due to all simulated collisions being deep in the gravity-dominated regime \citep{Housen1990}.

\subsection{Atmosphere Removal}

\citet{Denman2020} show that the likelihood of a collision removing all the atmosphere from a planet in one go is low, and if it does, it will be sufficiently energetic to cause catastrophic disruption.
The head-on collisions in this work are in strong agreement with their predictions for atmosphere loss, as shown in Figure \ref{fig:Res_HOfits}. 

Comparing our results for the mass remaining in the largest remnant for head-on collisions to those at higher impact parameters (given in Figure \ref{fig:Res_EHR}) we see that, for all cases where $b>0.3$, higher impact parameters mean a lower efficiency of atmosphere erosion. Considering that higher impact parameters are more likely than head-on (\citealt{Shoemaker1962} predict a most likely value of $45^{\circ}$), this means that a single impact removing the entirety of a Super-Earth's atmosphere is very unlikely. This implies that when we observe a Super-Earth mass planet with no substantial atmosphere it has likely either undergone multiple collisions or formed sufficiently close to its central star for photo-evaporation to have stripped the entire atmosphere.

In our resolution tests we observe an average difference of $7\%$ from our standard resolution for atmosphere mass (see appendix \ref{ap:ResolutionTesting}).  Our higher resolution simulations typically show slightly less atmosphere erosion than our lower resolution ones, so we would expect full atmosphere stripping in a real system to be less likely than in our simulations.

\subsection{Final Planet Radii}
\label{sec:Disc_Density}

\begin{figure}
 \includegraphics[width=\columnwidth]{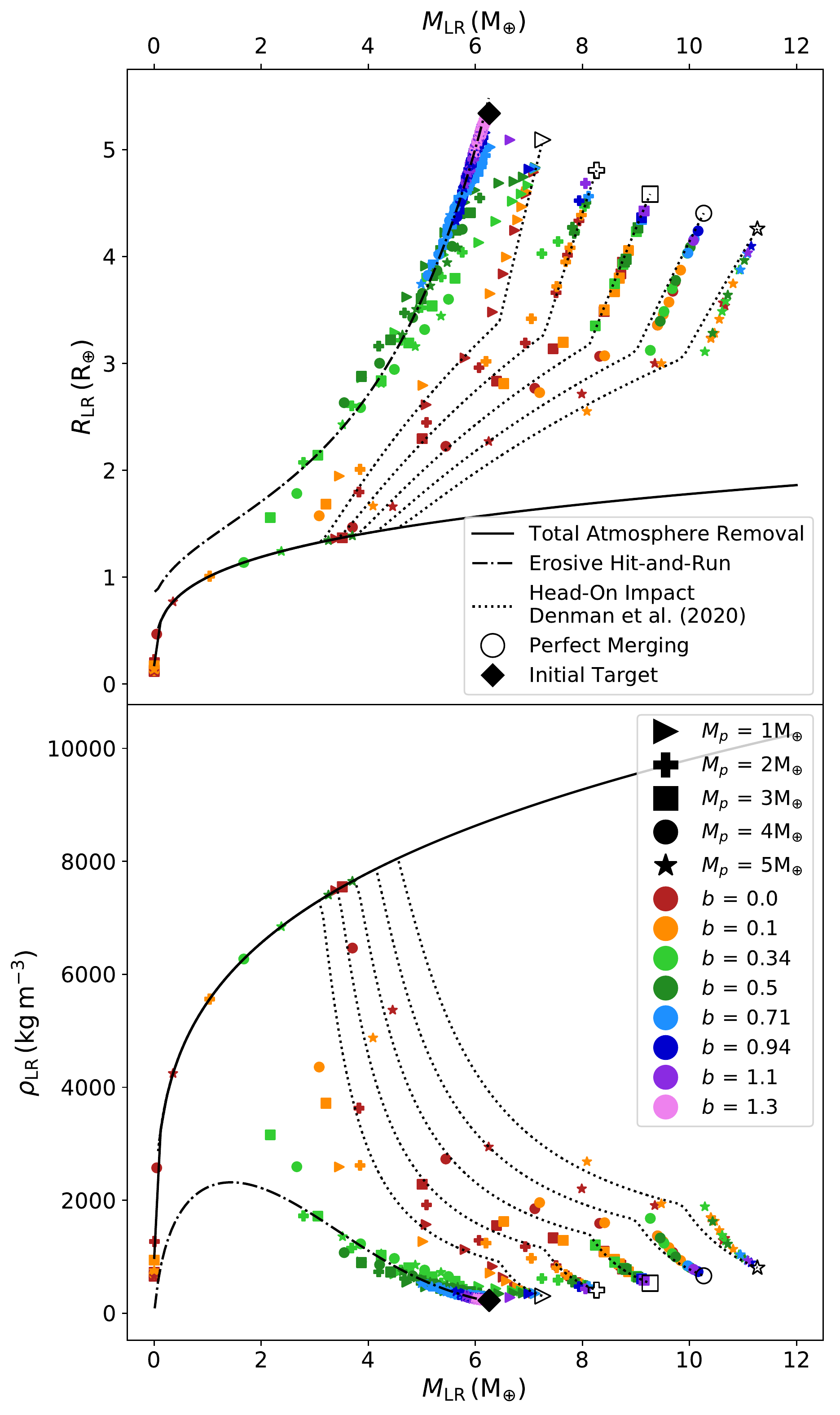} 
 \caption{
    \textit{Top:} The radii of the largest remnants generated by our collisions as a function of their masses. The colour of points indicates the impact parameter whereas the shape indicates the projectile mass. Lines indicate the expected results for total atmosphere erosion (solid line), and for partial atmosphere erosion in both the erosive hit-and-run (dash-dot) and head-on-like regimes \citep[dotted lines, from][]{Denman2020}. The open symbol at the end of each dotted line shows which projectile mass the line represents.  
	\textit{Bottom:} Corresponding densities as a function of largest post-collision remnant mass. Note that for real systems the upper limit for density is likely also dependent on material types in the core. Highly erosive collisions will strip mantle more readily than core, leaving a denser remnant than would be expected if the same ratio of core to mantle was maintained.
	}
 \label{fig:Disc_FinalRadii}
\end{figure}

As in \citet{Denman2020}, we have used \citet{Lopez2014}'s results to predict the final radii and densities of planets from their percentage composition post collision once all the bound material has settled back down, see Figure \ref{fig:Disc_FinalRadii}.

\citet{Lopez2014} model a planet's radius by splitting it into 3 contributions, these are: the core and mantle (modelled as a power law function of mass), the convective envelope (which is dependent on temperature, incident flux and planet age), and the radiative atmosphere (also dependent on temperature). For our results we used a stellar age of $5\,\mathrm{Gyr}$ (the most common age for nearby stars) and a flux of $100\,\mathrm{F_{\oplus}}$ (the expected flux of a planet orbiting at $0.1\,\mathrm{au}$ about a sun like star). The contribution of the radiative atmosphere is small for the regime we are dealing with ($\sim 0.1\,\mathrm{R_{\oplus}}$) so we ignore its contribution. 

The majority of higher impact parameter results in Figure \ref{fig:Disc_FinalRadii}  (greens, blues and purples) cluster in a single group, these are the erosive hit-and-run collisions, the dash-dotted line here is \citet{Lopez2014}'s model applied to the material composition fit of equation \ref{eq:Res_fH}. Lower impact parameter collisions on the other hand separate into multiple groups depending on their projectile mass (symbol shape); these groups agree with radius and density predictions using the \citet{Denman2020} prescription to predict material composition (dotted lines, with the shape of the open symbol at the end denoting projectile mass). The black line denotes the expected radius if the entirety of the atmosphere was eroded.

All simulated collisions result in ejection of atmosphere causing a reduction in planet radius typically leading to an increase in final density. Figure \ref{fig:Res_MaterialFractions} shows that the the atmosphere fraction for the majority of collisions clusters around a single function of total mass, as such so do the final radii of planets. The majority of collisions that are not close to this erosive hit-and-run prediction line instead follow the predictions for atmosphere mass fraction from \citet{Denman2020} for merging collisions.

Although from Figure \ref{fig:Disc_FinalRadii} it appears that these collisions could generate Earth analogues, i.e. objects with the Earth's mass and size, Figure \ref{fig:Res_MaterialFractions} reveals that to erode enough material to reach this mass these collisions eject a significant fraction of the mantle and as such they result in a much higher iron fraction than the Earth. The model we are using  to predict radius doesn't include core fraction for sake of simplicity, so we would expect these Earth mass planets to actually be smaller and denser. Considering the relationship between final iron content and final mass detailed in Figure \ref{fig:Res_MaterialFractions}, we expect the predictions here to underestimate density and overestimate radius by a factor that increases with decreasing final remnant mass.

Overall, Figure \ref{fig:Disc_FinalRadii} shows that a collision between a single target-projectile pair (these are each denoted by a different symbol) can result in a wide variety of radii and densities in the remnant planets. These simulations therefore provide direct support to the conclusions of \citet{Inamdar2016} that giant impacts can enrich the density diversity of Super-Earth planets.

\section{Conclusions}

Collisions between a Super-Earth sized target with an atmosphere, and a second Super-Earth sized projectile (with no atmosphere) tend to fall into two different collision regimes. At high impact parameter and velocity they have erosive hit-and-run collisions in which the target and projectile cores mostly remain intact and separate, but erosion and sometimes accretion occurs during the collision.
For low impact angles and velocities we instead observe the projectile and target combining into a single large remnant, following \citet{Denman2020}. Between these two regimes we observe graze-and-merge collisions where the target and projectile eventually merge into one large remnant, but not on the first collision.

We observe the transition between these two regimes to be well described by a simple escape velocity measure, the minimum velocity required for an erosive hit-and run being 
\begin{equation*}
  v_{\mathrm{split}}=\sqrt{\frac{2G(M_\mathrm{p}+M_\mathrm{t})}{b(R_\mathrm{p}+R_\mathrm{t})} }.  
\end{equation*}

For erosive hit-and-run collisions in which large secondary remnants arise, the dependence of the mass of the largest remnant on projectile mass is solely contained within the specific impact energy. In other words, all collisions in the erosive limit appear to follow the same pattern with regards to specific relative kinetic energy of impact regardless of projectile mass.

We have used our results to derive scaling laws for both the mass of atmosphere and also the total mass lost for both regimes for collisions within the parameter space tested. These scaling laws are summarised in appendix \ref{ap:Prescription} which details how they expand the prescription of \citet{Denman2020} to oblique impacts. 

The results of the simulations outlined in this paper reinforce the conclusion from \citet{Denman2020} that it is impossible to remove the entirety of the atmosphere of the target planet without removing a significant percentage of the mantle as well. Considering the higher probabilities of higher impact angles and lower velocity collisions, a giant impact being sufficiently energetic to remove the entirety of a planet's atmosphere is likely a rare occurrence.

Despite the low likelihood of removing the entire atmosphere of a planet with a single collision, the planets resulting from giant impacts cover a wide array of different radii and densities, reinforcing the hypothesis that planet-planet collisions are a cause of the large density diversity we observe in Super-Earth populations.

\section*{Acknowledgements}

This work was carried out using the computational facilities of the Advanced Computing Research Centre, University of Bristol - http://www.bristol.ac.uk/acrc/. TD acknowledges support from an STFC studentship (grant number: ST/R504634/1). ZML and PJC acknowledge support from UKRI. PJC acknowledges financial support from the Simons Foundation. We thank Christoph Mordasini for his work generating the atmosphere profiles. 

This research has made use of NASA’s Astrophysics Data System. We acknowledge the use of the {\sc python} libraries {\sc matplotlib} \citep{Hunter2007},  {\sc numpy} \citep{Harris2020}, and {\sc scipy} \citep{Virtanen2020}

\section*{Data Availability}

A summary of simulation results for this paper are provided in the appendices. Full simulation output is available from the authors on reasonable request. The modified version of Gadget-2 is available from \citet{Gadget2Planetary2022}. The equation of state tables for use with Gadget-2 are available from \citet{GADGET2EOS2019}.




\bibliographystyle{mnras}
\bibliography{references} 

\begin{thebibliography}{}
\makeatletter
\relax
\def\mn@urlcharsother{\let\do\@makeother \do\$\do\&\do\#\do\^\do\_\do\%\do\~}
\def\mn@doi{\begingroup\mn@urlcharsother \@ifnextchar [ {\mn@doi@}
  {\mn@doi@[]}}
\def\mn@doi@[#1]#2{\def\@tempa{#1}\ifx\@tempa\@empty \href
  {http://dx.doi.org/#2} {doi:#2}\else \href {http://dx.doi.org/#2} {#1}\fi
  \endgroup}
\def\mn@eprint#1#2{\mn@eprint@#1:#2::\@nil}
\def\mn@eprint@arXiv#1{\href {http://arxiv.org/abs/#1} {{\tt arXiv:#1}}}
\def\mn@eprint@dblp#1{\href {http://dblp.uni-trier.de/rec/bibtex/#1.xml}
  {dblp:#1}}
\def\mn@eprint@#1:#2:#3:#4\@nil{\def\@tempa {#1}\def\@tempb {#2}\def\@tempc
  {#3}\ifx \@tempc \@empty \let \@tempc \@tempb \let \@tempb \@tempa \fi \ifx
  \@tempb \@empty \def\@tempb {arXiv}\fi \@ifundefined
  {mn@eprint@\@tempb}{\@tempb:\@tempc}{\expandafter \expandafter \csname
  mn@eprint@\@tempb\endcsname \expandafter{\@tempc}}}

\bibitem[\protect\citeauthoryear{Alibert, Mordasini, Benz  \&
  Winisdoerffer}{Alibert et~al.}{2005}]{Alibert2005}
Alibert Y.,  Mordasini C.,  Benz W.,   Winisdoerffer C.,  2005, \mn@doi
  [Astronomy and Astrophysics] {10.1051/0004-6361:20042032}, 434, 343

\bibitem[\protect\citeauthoryear{Asphaug}{Asphaug}{2010}]{Asphaug2010a}
Asphaug E.,  2010, \mn@doi [Chemie der Erde] {10.1016/j.chemer.2010.01.004},
  70, 199

\bibitem[\protect\citeauthoryear{Asphaug}{Asphaug}{2014}]{Asphaug2014}
Asphaug E.,  2014, \mn@doi [Annual Review of Earth and Planetary Sciences]
  {10.1146/annurev-earth-050212-124057}, 42, 551

\bibitem[\protect\citeauthoryear{Barnes \& Raymond}{Barnes \&
  Raymond}{2004}]{Barnes2004}
Barnes R.,  Raymond S.~N.,  2004, \mn@doi [The Astrophysical Journal]
  {10.1086/423419}, 617, 569

\bibitem[\protect\citeauthoryear{Benz \& Asphaug}{Benz \&
  Asphaug}{1999}]{Benz1999}
Benz W.,  Asphaug E.,  1999, Icarus, 142, 5

\bibitem[\protect\citeauthoryear{Bonomo et~al.,}{Bonomo
  et~al.}{2019}]{Bonomo2019}
Bonomo A.~S.,  et~al., 2019, \mn@doi [Nature Astronomy]
  {10.1038/s41550-018-0684-9}, 3, 416

\bibitem[\protect\citeauthoryear{Carter}{Carter}{2022}]{Gadget2Planetary2022}
Carter P.,  2022, {PhilJCarter/gadget2-planetary: v1.0.0: Gadget2-Planetary
  initial versioning release}, \mn@doi{10.5281/ZENODO.5879324}, \url
  {https://doi.org/10.5281/zenodo.5879324#.YfFfQVFmPQ9.mendeley}

\bibitem[\protect\citeauthoryear{Carter, Lock  \& Stewart}{Carter
  et~al.}{2019}]{GADGET2EOS2019}
Carter P.~J.,  Lock S.~J.,   Stewart S.~T.,  2019, {Replication Data for: ``The
  energy budgets of giant impacts''}, \mn@doi{10.7910/DVN/YYNJSX}, \url
  {https://doi.org/10.7910/DVN/YYNJSX}

\bibitem[\protect\citeauthoryear{{\'{C}}uk \& Stewart}{{\'{C}}uk \&
  Stewart}{2012}]{Matija2012}
{\'{C}}uk M.,  Stewart S.~T.,  2012, \mn@doi [Science]
  {10.1103/PhysRevB.93.235445}, pp 1047--1053

\bibitem[\protect\citeauthoryear{Denman, Leinhardt, Carter  \&
  Mordasini}{Denman et~al.}{2020}]{Denman2020}
Denman T.~R.,  Leinhardt Z.~M.,  Carter P.~J.,   Mordasini C.,  2020, \mn@doi
  [Monthly Notices of the Royal Astronomical Society] {10.1093/MNRAS/STAA1623},
  496, 1166

\bibitem[\protect\citeauthoryear{Emsenhuber \& Asphaug}{Emsenhuber \&
  Asphaug}{2019}]{Emsenhuber2019a}
Emsenhuber A.,  Asphaug E.,  2019, \mn@doi [The Astrophysical Journal]
  {10.3847/1538-4357/ab2f8e}, 881, 102

\bibitem[\protect\citeauthoryear{Fabrycky et~al.,}{Fabrycky
  et~al.}{2014}]{Fabrycky2014}
Fabrycky D.~C.,  et~al., 2014, \mn@doi [Astrophysical Journal]
  {10.1088/0004-637X/790/2/146}, 790

\bibitem[\protect\citeauthoryear{Fang \& Margot}{Fang \&
  Margot}{2013}]{Fang2013}
Fang J.,  Margot J.~L.,  2013, \mn@doi [Astrophysical Journal]
  {10.1088/0004-637X/767/2/115}, 767, 115

\bibitem[\protect\citeauthoryear{Gabriel, Jackson, Asphaug, Reufer, Jutzi  \&
  Benz}{Gabriel et~al.}{2020}]{Gabriel2020}
Gabriel T. S.~J.,  Jackson A.~P.,  Asphaug E.,  Reufer A.,  Jutzi M.,   Benz
  W.,  2020, \mn@doi [The Astrophysical Journal] {10.3847/1538-4357/ab528d},
  892, 40

\bibitem[\protect\citeauthoryear{Gardiner}{Gardiner}{2015}]{BlueCrystal3}
Gardiner C.,  2015, {Blue Crystal Phase 3}, \url
  {https://www.acrc.bris.ac.uk/acrc/phase3.htm}

\bibitem[\protect\citeauthoryear{Gardiner}{Gardiner}{2017}]{BlueCrystal4}
Gardiner C.,  2017, {Blue Crystal Phase 4}, \url
  {https://www.acrc.bris.ac.uk/acrc/phase4.htm}

\bibitem[\protect\citeauthoryear{Genda, Kokubo  \& Ida}{Genda
  et~al.}{2012}]{Genda2012}
Genda H.,  Kokubo E.,   Ida S.,  2012, \mn@doi [Astrophysical Journal]
  {10.1088/0004-637X/744/2/137}, 744, 4

\bibitem[\protect\citeauthoryear{Genda, Fujita, Kobayashi, Tanaka  \&
  Abe}{Genda et~al.}{2015}]{Genda2015}
Genda H.,  Fujita T.,  Kobayashi H.,  Tanaka H.,   Abe Y.,  2015, \mn@doi
  [Icarus] {10.1016/j.icarus.2015.08.029}, 262, 58

\bibitem[\protect\citeauthoryear{Gillion et~al.,}{Gillion
  et~al.}{2017}]{Gillon2017}
Gillion M.,  et~al., 2017, \mn@doi [Nature] {10.1038/nature21360}, 542, 456

\bibitem[\protect\citeauthoryear{Harris et~al.,}{Harris
  et~al.}{2020}]{Harris2020}
Harris C.~R.,  et~al., 2020, \mn@doi [Nature] {10.1038/s41586-020-2649-2}, 585,
  357

\bibitem[\protect\citeauthoryear{Hartmann \& Davis}{Hartmann \&
  Davis}{1975}]{Hartmann1975}
Hartmann W.~K.,  Davis D.~R.,  1975, \mn@doi [Icarus]
  {10.1016/0019-1035(75)90070-6}, 24, 504

\bibitem[\protect\citeauthoryear{Housen \& Holsapple}{Housen \&
  Holsapple}{1990}]{Housen1990}
Housen K.~R.,  Holsapple K.~A.,  1990, \mn@doi [Icarus]
  {10.1016/0019-1035(90)90168-9}, 84, 226

\bibitem[\protect\citeauthoryear{Hunter}{Hunter}{2007}]{Hunter2007}
Hunter J.~D.,  2007, \mn@doi [Computing in Science {\&} Engineering]
  {10.1109/MCSE.2007.55}, 9, 90

\bibitem[\protect\citeauthoryear{Hwang, Chatterjee, Lombardi, Steffen  \&
  Rasio}{Hwang et~al.}{2017}]{Hwang2017b}
Hwang J.,  Chatterjee S.,  Lombardi J.,  Steffen J.,   Rasio F.,  2017, \mn@doi
  [The Astrophysical Journal] {10.3847/1538-4357/aa9d42}, 852, 41

\bibitem[\protect\citeauthoryear{Inamdar \& Schlichting}{Inamdar \&
  Schlichting}{2015}]{Inamdar2015}
Inamdar N.~K.,  Schlichting H.~E.,  2015, \mn@doi [Monthly Notices of the Royal
  Astronomical Society] {10.1093/mnras/stv030}, 448, 1751

\bibitem[\protect\citeauthoryear{Inamdar \& Schlichting}{Inamdar \&
  Schlichting}{2016}]{Inamdar2016}
Inamdar N.~K.,  Schlichting H.~E.,  2016, \mn@doi [The Astrophysical Journal]
  {10.3847/2041-8205/817/2/l13}, 817, L13

\bibitem[\protect\citeauthoryear{Kegerreis et~al.,}{Kegerreis
  et~al.}{2018}]{Kegerreis2018}
Kegerreis J.~A.,  et~al., 2018, \mn@doi [The Astrophysical Journal]
  {10.3847/1538-4357/aac725}, 861, 52

\bibitem[\protect\citeauthoryear{Kegerreis, Eke, Catling, Massey, Teodoro  \&
  Zahnle}{Kegerreis et~al.}{2020}]{Kegerreis2020}
Kegerreis J.~A.,  Eke V.~R.,  Catling D.~C.,  Massey R.~J.,  Teodoro L. F.~A.,
   Zahnle K.~J.,  2020, \mn@doi [The Astrophysical Journal]
  {10.3847/2041-8213/abb5fb}, 901, L31

\bibitem[\protect\citeauthoryear{Leinhardt \& Stewart}{Leinhardt \&
  Stewart}{2012}]{Leinhardt2012}
Leinhardt Z.~M.,  Stewart S.~T.,  2012, \mn@doi [Astrophysical Journal]
  {10.1088/0004-637X/745/1/79}, 745, 79

\bibitem[\protect\citeauthoryear{Liu, Hori, Lin  \& Asphaug}{Liu
  et~al.}{2015}]{Liu2015}
Liu S.~F.,  Hori Y.,  Lin D.~N.,   Asphaug E.,  2015, \mn@doi [Astrophysical
  Journal] {10.1088/0004-637X/812/2/164}, 812, 164

\bibitem[\protect\citeauthoryear{Lock, Stewart, Petaev, Leinhardt, Mace,
  Jacobsen  \& Cuk}{Lock et~al.}{2018}]{Lock2018}
Lock S.~J.,  Stewart S.~T.,  Petaev M.~I.,  Leinhardt Z.,  Mace M.~T.,
  Jacobsen S.~B.,   Cuk M.,  2018, \mn@doi [Journal of Geophysical Research:
  Planets] {10.1002/2017JE005333}, 123, 910

\bibitem[\protect\citeauthoryear{Lopez \& Fortney}{Lopez \&
  Fortney}{2014}]{Lopez2014}
Lopez E.~D.,  Fortney J.~J.,  2014, \mn@doi [Astrophysical Journal]
  {10.1088/0004-637X/792/1/1}, 792, 1

\bibitem[\protect\citeauthoryear{Marcus, Stewart, Sasselov  \&
  Hernquist}{Marcus et~al.}{2009}]{Marcus2009}
Marcus R.~A.,  Stewart S.~T.,  Sasselov D.,   Hernquist L.,  2009, \mn@doi
  [Astrophysical Journal] {10.1088/0004-637X/700/2/L118}, 700, 118

\bibitem[\protect\citeauthoryear{Marcus, Sasselov, Stewart  \&
  Hernquist}{Marcus et~al.}{2010}]{Marcus2010}
Marcus R.~A.,  Sasselov D.,  Stewart S.~T.,   Hernquist L.,  2010, \mn@doi
  [Astrophysical Journal Letters] {10.1088/2041-8205/719/1/L45}, 719, 45

\bibitem[\protect\citeauthoryear{Melosh \& Vickery}{Melosh \&
  Vickery}{1989}]{Melosh1989}
Melosh H.~J.,  Vickery A.~M.,  1989, \mn@doi [Nature] {10.1038/338487a0}, 338,
  487

\bibitem[\protect\citeauthoryear{Mills, Fabrycky, Migaszewski, Ford, Petigura
  \& Isaacson}{Mills et~al.}{2016}]{Mills2016}
Mills S.~M.,  Fabrycky D.~C.,  Migaszewski C.,  Ford E.~B.,  Petigura E.,
  Isaacson H.,  2016, \mn@doi [Nature] {10.1038/nature17445}, 533, 509

\bibitem[\protect\citeauthoryear{Mordasini}{Mordasini}{2018}]{Mordasini2018}
Mordasini C.,  2018, in , Handbook of Exoplanets.
Springer International Publishing, pp 2425--2474,
  \mn@doi{10.1007/978-3-319-55333-7{\_}143}

\bibitem[\protect\citeauthoryear{Movshovitz, Nimmo, Korycansky, Asphaug  \&
  Owen}{Movshovitz et~al.}{2016}]{Movshovitz2016}
Movshovitz N.,  Nimmo F.,  Korycansky D.~G.,  Asphaug E.,   Owen J.~M.,  2016,
  \mn@doi [Icarus] {10.1016/j.icarus.2016.04.018}, 275, 85

\bibitem[\protect\citeauthoryear{Raymond \& Morbidelli}{Raymond \&
  Morbidelli}{2020}]{Raymond2020}
Raymond S.~N.,  Morbidelli A.,  2020, {Planet formation: key mechanisms and
  global models}, \url {http://arxiv.org/abs/2002.05756}

\bibitem[\protect\citeauthoryear{Rogers, Bodenheimer, Lissauer  \&
  Seager}{Rogers et~al.}{2011}]{Rogers2011}
Rogers L.~A.,  Bodenheimer P.,  Lissauer J.~J.,   Seager S.,  2011, \mn@doi
  [Astrophysical Journal] {10.1088/0004-637X/738/1/59}, 738

\bibitem[\protect\citeauthoryear{Shoemaker \& Hackman}{Shoemaker \&
  Hackman}{1962}]{Shoemaker1962}
Shoemaker E.~M.,  Hackman R.~J.,  1962, \mn@doi [Symposium - International
  Astronomical Union] {10.1017/s007418090017826x}

\bibitem[\protect\citeauthoryear{Springel}{Springel}{2005}]{Springel2005}
Springel V.,  2005, \mn@doi [Monthly Notices of the Royal Astronomical Society]
  {10.1111/j.1365-2966.2005.09655.x}, 364, 1105

\bibitem[\protect\citeauthoryear{Stewart \& Leinhardt}{Stewart \&
  Leinhardt}{2009}]{Stewart2009}
Stewart S.~T.,  Leinhardt Z.~M.,  2009, \mn@doi [Astrophysical Journal]
  {10.1088/0004-637X/691/2/L133}, 691, L133

\bibitem[\protect\citeauthoryear{Valencia, O'Connell  \& Sasselov}{Valencia
  et~al.}{2006}]{Valencia2006}
Valencia D.,  O'Connell R.~J.,   Sasselov D.,  2006, \mn@doi [Icarus]
  {10.1016/j.icarus.2005.11.021}, 181, 545

\bibitem[\protect\citeauthoryear{Virtanen et~al.,}{Virtanen
  et~al.}{2020}]{Virtanen2020}
Virtanen P.,  et~al., 2020, \mn@doi [Nature Methods]
  {10.1038/s41592-019-0686-2}, 17, 261

\bibitem[\protect\citeauthoryear{Volk \& Gladman}{Volk \&
  Gladman}{2015}]{Volk2015}
Volk K.,  Gladman B.,  2015, \mn@doi [Astrophysical Journal Letters]
  {10.1088/2041-8205/806/2/L26}, 806, L26

\bibitem[\protect\citeauthoryear{Ward}{Ward}{1986}]{Ward1986a}
Ward W.~R.,  1986, \mn@doi [Icarus] {10.1016/0019-1035(86)90182-X}, 67, 164

\bibitem[\protect\citeauthoryear{Watt, Leinhardt  \& Su}{Watt
  et~al.}{2021}]{Watt2021}
Watt L.,  Leinhardt Z.,   Su K. Y.~L.,  2021, \mn@doi [Monthly Notices of the
  Royal Astronomical Society] {10.1093/mnras/stab106}, 502, 2984

\makeatother
\end{thebibliography}




\appendix
\section{Resolution Testing}
\label{ap:ResolutionTesting}
\begin{figure}
 \includegraphics[width=\columnwidth]{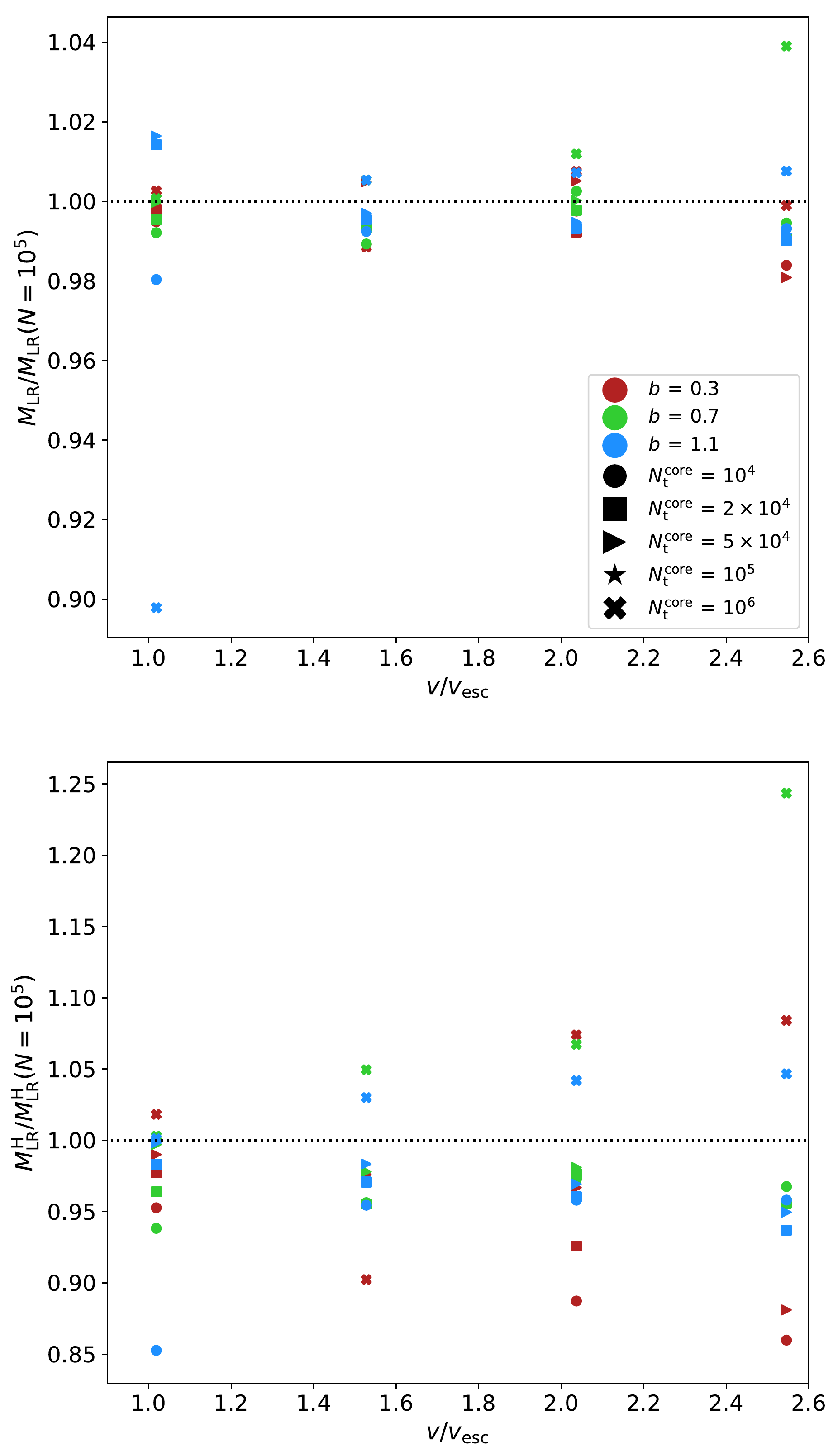}
 \caption{\textit{Top:} The ratio of largest remnant mass for various simulation resolutions to the remnant mass for our standard resolution ($N=10^5$) compared to impact velocity. The majority of runs here give results within $2\%$ of the standard resolution chosen.
 \textit{Bottom:} Atmosphere mass remaining as a fraction of the atmosphere mass obtained at our standard resolution at different impact velocities. Due to the lower resolution of the atmosphere compared to the total mass this shows less close agreement, the majority of results still fall within $6\%$ of the values from the chosen resolution.
 }
 \label{fig:NM_ResTest}
\end{figure}

Many previous papers, e.g. \citet{Genda2015}, have focused on the effect of resolution on the results of SPH simulations, despite this the precise dependence of loss overestimation is unknown. We provide here the results of an array of different resolution simulations to measure these effects for our collisions. The resolutions used for these tests were $10^4, 2\times10^4, 5\times10^4$ and $10^6$ particles in the $5\mathrm{M_{\oplus}}$ target core and mantle compared to the standard resolution of $10^5$ particles in the target core and mantle. 

A comparison of the results for these different resolutions is given in Figure \ref{fig:NM_ResTest}. For mass of the largest remnant our results show agreement between the resolutions to within $2\%$ for all but two tests. For atmosphere mass remaining in the largest remnant the results detailed in Figure \ref{fig:NM_ResTest} show a greater spread between resolutions; the majority of results agree to within $7\%$, however there are some significant outliers which differ by up to $25\%$.

We also observe in our results that the amount of atmosphere remaining seems slightly dependent on resolution, with the higher resolution simulations typically retaining more atmosphere after an impact. What is likely happening is that the greater spatial resolution for the higher resolution simulations means that atmosphere particles interact more frequently and there is thus more efficient energy partitioning between particles.

Overall these resolution tests show a good agreement with our standard resolution results both mass of the largest remnant and for atmosphere mass.

\section{Erosive Hit-and-run fitting process}
\label{ap:EHR_Construction}

\begin{figure*}
    \centering
    \includegraphics[width=1.65\columnwidth]{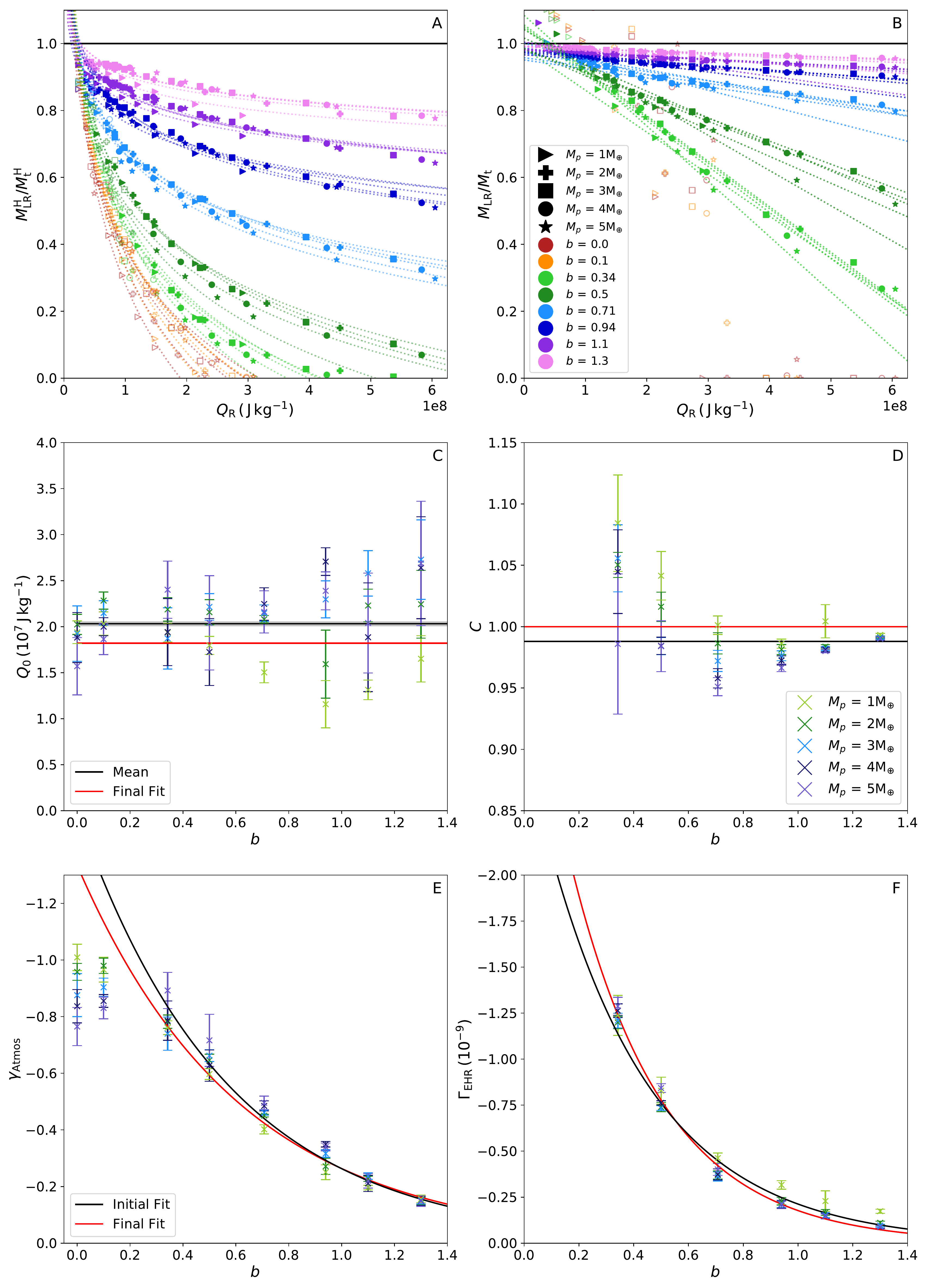}
    \caption{ 
    \textit{A:} Atmosphere mass in the largest remnant as a function of $Q_\mathrm{R}$; this has been fit for each impact parameter and projectile mass by a series of power law fits (dotted lines). Erosive hit-and-run collisions are given by filled symbols, all other collisions are open symbols.
    \textit{B:} Largest remnant mass normalised by target mass compared to $Q_\mathrm{R}$. The erosive hit-and-run collisions for impact parameter and projectile mass have been fit with linear fits.  
    \textit{C:} Minimum energy required for atmosphere to be ejected, $Q_\mathrm{0}$, for each power law fit in A as a function of impact parameter. There is some scatter with no obvious physical cause so the final fit (red line) uses a common value for all impact angles and projectiles. The mean $Q_\mathrm{0}$ is shown in black.
    \textit{D:} The intercept of the linear fits to largest remnant mass versus impact energy as a function of impact parameter. Many of these intercepts fall below the target mass (the mean value is in black), implying that mass could be lost without input energy, this is un-physical so we set the intercept to the target mass (unity) for all fits.
    \textit{E:} The coefficient for power law fits to atmosphere mass in the largest remnant versus impact energy, this itself follows a power law function of the impact parameter. An initial power law fit to these coefficients is shown in black. The final fit from figure \ref{fig:Res_EHR}, which fits for impact parameter and impact energy dependence at the same time, is shown in red.
    \textit{F:} The gradient of linear fits to projectile mass in the largest remnant versus impact energy fixed to go through zero mass loss for zero impact energy. These gradients follow a power law (black line).
    }
    \label{fig:ap_EHRconst}
\end{figure*}
This section summarises the reasoning behind the functional forms of the final fits used for the total mass and atmosphere masses of the largest remnant in sections \ref{sec:Res_EHR} and \ref{sec:Res_atmosEHR}.

\subsection{Atmosphere fitting}

The left side of Figure \ref{fig:ap_EHRconst} shows atmosphere mass in the largest remnant as a function of specific impact energy and the dependence of fit parameters on the impact parameter, $b$. The top left panel shows atmosphere mass normalised by initial target atmosphere mass compared to impact energy, $Q_{\mathrm{R}}$. Data for each individual projectile mass and impact parameter are fit with the following power law: \begin{equation}
    \frac{ M^{\mathrm{H}}_{\mathrm{LR}} }{M^{\mathrm{H}}_{\mathrm{t}} } = \gamma_{\mathrm{atmos}} \log_{10}\left(Q_{\mathrm{R}}\right)+ c_{\mathrm{atmos}},
\end{equation}
where $\gamma_{\mathrm{atmos}}$ and $c_{\mathrm{atmos}}$ are fit parameters.

Each of these fits (dotted lines in the top left graph of Figure \ref{fig:ap_EHRconst}) cross the target atmosphere mass at an impact energy of approximately $2\times10^{7}\,\mathrm{J\,kg^{-1}}$. The exact value of this collision energy is given by 
\begin{equation}
    Q_{\mathrm{0}} = 10^{ \frac{1-c_{\mathrm{atmos}} }{\gamma{\mathrm{atmos}}} },
\end{equation}
for each fit. This value has a reasonably low amount of scatter, that does not appear to show any impact parameter or projectile mass dependence, so we elected to use an impact parameter dependant fit parameter for the final combined fit (section \ref{sec:Res_atmosEHR}. The $Q_\mathrm{0}$ measured in the final fit is given by the red line in the left middle graph in Figure \ref{fig:ap_EHRconst}, the mean value of $Q_{\mathrm{0}}$ from our separate fits for each proectile mass and impact parameter is given by the black line.

The power law coefficient $\gamma_{\mathrm{atmos}}$ for each of these individual fits is given in the bottom left graph of Figure \ref{fig:ap_EHRconst}. Apart from the head-on like collisions ($b<0.2$), these have a power law dependence on impact parameter; the black line shows a power law fit to these coefficients, whereas the red line is the impact parameter dependence of the coefficient obtained for our final fit (equation \ref{eq:Res_LRAtmosphereFit}) where we fit all the projectiles and impact angles simultaneously.

\subsection{Largest Remnant mass fitting}

The top right graph of Figure \ref{fig:ap_EHRconst} compares largest remnant mass and impact energy, the erosive hit-and-run collisions are shown with solid symbols. These erosive hit-and-run collisions show a linear dependence of remnant mass on impact energy, each separate impact parameter and projectile mass was therefore fit with:
\begin{equation}
    \frac{ M_{\mathrm{LR}} }{ M_{\mathrm{t}} } = \Gamma_{\mathrm{EHR}}Q_{\mathrm{R}}+C,
\end{equation}
where $\Gamma_{\mathrm{EHR}}$ and $C$ are the gradient and the intercept respectively.

The intercept for each of these fits is shown in the middle right graph of Figure \ref{fig:ap_EHRconst}. Many of these values are below the target mass (less than 1), as is the mean intercept (the black line), this is unphysical as it implies material can be lost from the target without any impact energy. What we actually observe at low impact energies is a curve upwards away from the linear fit, so it is likely that mass loss for low impact energies, close to the transition region, is non-linear. However, we chose to use the simplest possible model to describe the data and so used a linear fit fixed to the target mass for zero impact energy (red line), i.e.\ $C=1$, despite the low energy differences this shows good agreement with the majority of the data.

The bottom graph in Figure \ref{fig:ap_EHRconst} details the dependence of the gradient of the largest remnant mass to impact energy relation on impact parameter. The gradient becomes less negative by a decreasing amount with increasing impact parameter, so we fit it with a power law (black line) which shows good agreement. The dependence of gradient on impact parameter obtained from our final fit (equation \ref{eq:MLR_Equation}), which simultaneously fit to all projectile masses and impact parameters, is shown in red.

\section{Graze-and-Merge re-collision time}
\label{ap:Graze-and-Merge}

To approximate the longest possible orbit time that could occur before re-collision, without the secondary remnant escaping the largest remnant's gravitational influence, we first need to calculate the Hill radius of the target planet $r_\mathrm{H}$:
\begin{equation}
    r_{\mathrm{H}}=a_{\mathrm{i}}\sqrt[\leftroot{-2}\uproot{10}3]{\frac{M_\mathrm{t}}{3M_\mathrm{0}}},
\end{equation}
where $a_{\mathrm{i}}$ is the semi-major axis of the target about the star it orbits and $M_{\mathrm{0}}$ that star's mass.
The semi-major axis of an orbit about the target after a grazing collision is half the sum of the apoapsis and periapsis distance and so the longest possible orbit of the secondary remnant without it escaping is  $a_\mathrm{ii}=(r_\mathrm{H}+R_\mathrm{t}+R_\mathrm{p})/2$. From this semi-major axis we can calculate the orbital period of this limiting orbit of the projectile about the target to be 
\begin{equation}
T=2\pi\sqrt{\frac{a^3_{\mathrm{ii}}}{G(M_\mathrm{t}+M_\mathrm{p})}}. 
\end{equation}
Applying this to our scenario, for a close orbiting super Earth $0.1\,\mathrm{au}$ from a sun-like star (like we use in section \ref{sec:Disc_Density} for density estimates) this gives us re-collision times of $5-7\,\mathrm{hrs}$ for Super-Earth mass projectiles. This time period is well within the time limit of $27.7\,\mathrm{hrs}$ of our simulations. So we would be able to model such re-collisions in our simulations. We do not model the central star's gravity in our simulations, however, so our simulations include graze-and-merge collisions on longer re-collision timescales, as such we removed the graze-and-merge collisions from our collision fits so that they could remain agnostic to orbital distance.

\section{Remeasuring the point of collision}
\label{ap:PointOfCollision}

\begin{figure}
 \includegraphics[width=0.85\columnwidth]{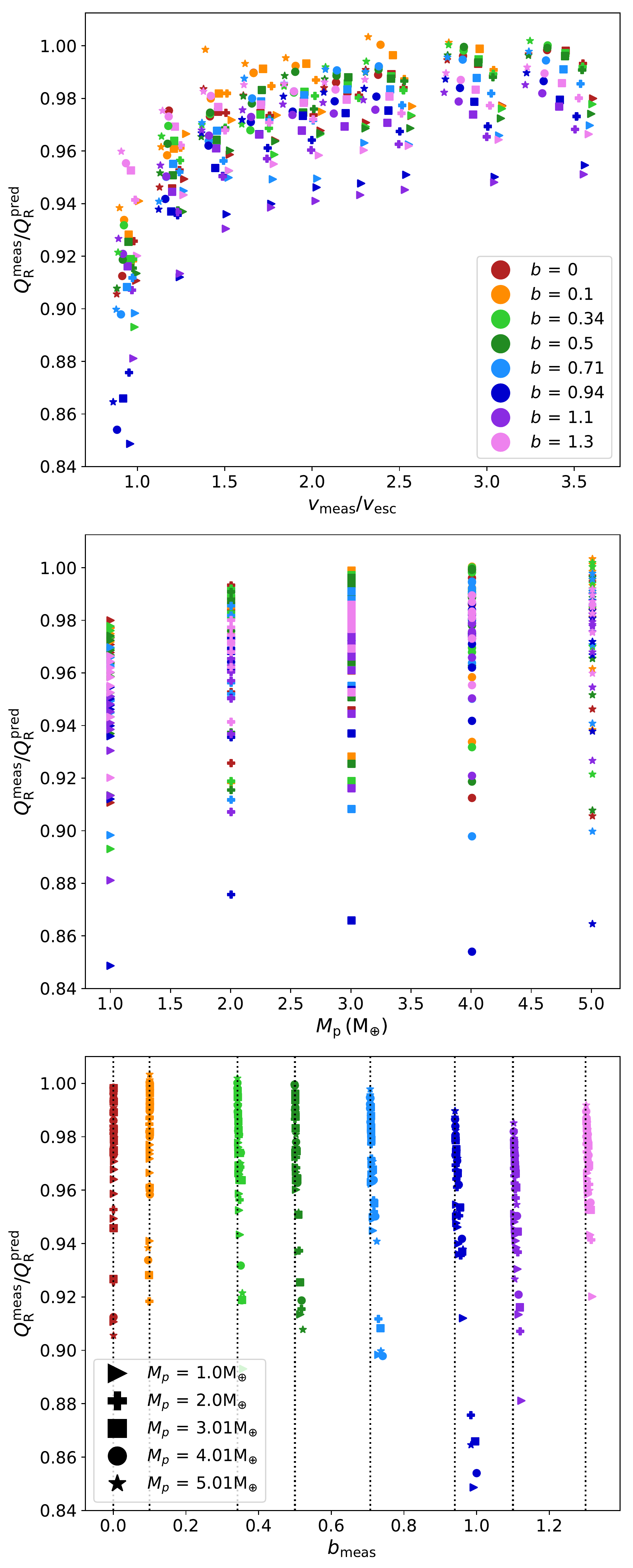}
 \caption{
    Variation of the specific impact energy as a function of other collision parameters. The colour of the points indicates the impact parameter, the shape indicates the mass of the projectile.
    \textit{Top:} Ratio of measured to expected $Q_\mathrm{R}$ vs measured impact velocity. There is a significant decrease in energy compared to the prediction for low velocities where a change in momentum due to drag can cause a much larger proportional change in velocity and thus kinetic energy.
	\textit{Middle:} Projectile mass compared with measured to expected $Q_{\mathrm{R}}$ ratio. Although there is a slight decrease in the mean impact energy for smaller projectiles due to lower momenta, projectile mass does not have a significant effect. 
	\textit{Bottom:} The measured to expected $Q_\mathrm{R}$ ratio compared with measured impact parameter. Impact parameter difference does not appear to cause a significant change to impact energy; note, however, that impacts with the largest change in kinetic energy tend to show the largest change from expected impact parameter (shown with dotted lines).
	}
 \label{fig:disc_PontOfImpact}
\end{figure}
  
 For ease of comparison to previous works on cratering and giant impacts (for example \citealt{Leinhardt2012,Marcus2009,Marcus2010}) and also to give a simple, reproducible approach that is independent of atmosphere scale height, we have chosen to measure our impact parameters at the point of collision between the mantle surfaces of projectile and target, or at the point of closest approach for highly grazing collisions. This method is not without its drawbacks however, the main one which needs to be considered is the effect of atmospheric drag.

To understand the difference caused by this drag we re-simulate the point of collision at a high snapshot output frequency and remeasure the relative velocity, separation, and impact parameter of the centre of masses at the point where the mantles collide (or reach their point of closest approach if $b>1$).

Because of this atmospheric drag, the velocity at the point of collision for an impact is reduced slightly. This means that the measured specific relative kinetic energy of impact, $Q_{\mathrm{R}}^{\mathrm{meas}}$, is typically less than that initially predicted $Q_{\mathrm{R}}^{\mathrm{pred}}$. Figure \ref{fig:disc_PontOfImpact} compares the ratio of measured to expected $Q_{\mathrm{R}}$ with  various collision parameters it is likely to be dependent on. The top two graphs compare the energy ratio to relative velocity and projectile mass; as might be expected, lower velocities and projectile masses typically mean a greater difference in impact energy, because the change in momentum due to the drag force is a larger percentage of the overall momentum of the projectile. The decrease we see in measured $Q_\mathrm{R}$ for lower impact velocities is stronger than that for lower projectile masses, this is due to $Q_{R}$ having a squared dependence on velocity (equation \ref{eq:Q_R}). The dependence of $Q_{R}$ on projectile mass is somewhat more complicated, being somewhere between approximately zero proportionality for $M_{\mathrm{p}} \approx M_{\mathrm{t}}$ and approximately linear dependence for $M_{\mathrm{p}} << M_\mathrm{t}$.
 
 Predicted impact parameter does not appear to have a significant effect on the amount of kinetic energy lost to atmospheric drag (see bottom panel Figure \ref{fig:disc_PontOfImpact}). We do, however, observe that, especially at higher impact parameters, the greater the measured deflection from the intended impact parameter the larger the percentage decrease in impact energy. Atmospheric pressure is deflecting the projectiles away from the target causing them to spend more time passing through the atmosphere before collision, thus producing a greater decrease in projectile momentum and therefore impact energy.
 
 For the majority of collisions, apart from those at the lowest impact velocity, the drag effects on kinetic energy are relatively small, typically causing a decrease of $<6\%$. Since this effect is small we therefore used the expected parameters for our analyses in this paper.

\section{Data tables}
\label{ap:DataTables}
We present here data tables, Table \ref{tab:PlanetData} provides details for each of the bodies we simulated collisions with, the other Tables each list the results of our simulations for a particular projectile mass. All collisions used a $6.25\,\mathrm{M_\oplus}$ target, the projectile mass for each of the collision results tables are as follows: Table \ref{tab:CollisionData_1} is for $1\,\mathrm{M_\oplus}$ projectiles, Table \ref{tab:CollisionData_2} is for $2\,\mathrm{M_\oplus}$ projectiles, Table \ref{tab:CollisionData_3} is for $3\,\mathrm{M_\oplus}$ projectiles, Table \ref{tab:CollisionData_4} is for $4\,\mathrm{M_\oplus}$ projectiles, and Table \ref{tab:CollisionData_5} is for $5\,\mathrm{M_\oplus}$ projectiles.

\begin{table}
\caption{Initial masses, $M$ of each planet used in our simulations, as well as the masses of each component (Fe -- iron, Si -- forsterite, H -- hydrogen) in each body, and the radius of the outermost simulation particle in each material layer. T indicates the planet was used as a target, whereas P were projectiles. The final column details the measured gravitational binding energy of each object. }
\label{tab:PlanetData}
\begin{tabular}{cccccccccc}
\hline
 Type & $M$ & $M_{\mathrm{Fe}}$ & $R_{\mathrm{Fe}}$ & $M_{\mathrm{Si}}$ & $R_{\mathrm{Si}}$ & $M_{\mathrm{H}}$ & $R_{\mathrm{H}}$ & $U_{\mathrm{g}}^{\mathrm{n}}$\\ 
     & $\mathrm{M_{\oplus}}$ & $\mathrm{M_{\oplus}}$ & $\mathrm{R_{\oplus}}$ & $\mathrm{M_{\oplus}}$ & $\mathrm{R_{\oplus}}$ & $\mathrm{M_{\oplus}}$ & $\mathrm{R_{\oplus}}$ & $\mathrm{J}$ \\ 

\hline
T & 6.26 & 1.66 & 0.76 & 3.34 & 1.51 & 1.25 & 5.34 & 5.06$\times$10$^{33}$ \\ 
P & 1.00 & 0.33 & 0.50 & 0.67 & 0.96 & 0.00 & - & 2.53$\times$10$^{32}$ \\ 
P & 2.00 & 0.67 & 0.60 & 1.33 & 1.17 & 0.00 & - & 8.40$\times$10$^{32}$ \\ 
P & 3.01 & 1.01 & 0.68 & 1.99 & 1.32 & 0.00 & - & 1.70$\times$10$^{33}$ \\ 
P & 4.01 & 1.32 & 0.72 & 2.69 & 1.43 & 0.00 & - & 2.79$\times$10$^{33}$ \\ 
P & 5.01 & 1.66 & 0.77 & 3.34 & 1.52 & 0.00 & - & 4.13$\times$10$^{33}$ \\ 

\hline
\end{tabular}
\end{table}

\begin{table}
\caption{A summary of the results of the simulations of collisions of the $1\,\mathrm{M_\oplus}$ projectile with the $6.25\,\mathrm{M_\oplus}$ target. Full planet data is given in Table \ref{tab:PlanetData}, the rest of the collision results are in Tables \ref{tab:CollisionData_2}, \ref{tab:CollisionData_3}, \ref{tab:CollisionData_4}, and \ref{tab:CollisionData_5}. }
\label{tab:CollisionData_1}
\begin{tabular}{cccccccccc}
\hline
$b$ & $v$ & $M_{\mathrm{LR}}$ & $f_{\mathrm{Fe}}$ & $f_{\mathrm{Si}}$ & $f_{\mathrm{H}}$ & $M_{\mathrm{SLR}}$ \\ 
 & $\mathrm{km\,s^{-1}}$ & $\mathrm{M_{\oplus}}$ & & & & $\mathrm{M_{\oplus}}$\\ 
\hline
0.00 & 20.0 & 7.09 & 0.28 & 0.57 & 0.15 & - \\ 
0.00 & 25.0 & 6.95 & 0.29 & 0.58 & 0.14 & - \\ 
0.00 & 30.0 & 6.73 & 0.30 & 0.59 & 0.11 & - \\ 
0.00 & 35.0 & 6.52 & 0.31 & 0.61 & 0.08 & - \\ 
0.00 & 40.0 & 6.32 & 0.32 & 0.62 & 0.06 & - \\ 
0.00 & 45.0 & 5.81 & 0.34 & 0.62 & 0.04 & - \\ 
0.00 & 50.0 & 5.08 & 0.39 & 0.58 & 0.02 & - \\ 
0.10 & 20.0 & 7.11 & 0.28 & 0.56 & 0.15 & - \\ 
0.10 & 25.0 & 6.96 & 0.29 & 0.58 & 0.14 & - \\ 
0.10 & 27.5 & 6.87 & 0.29 & 0.58 & 0.13 & - \\ 
0.10 & 30.0 & 6.79 & 0.29 & 0.59 & 0.12 & - \\ 
0.10 & 35.0 & 6.58 & 0.30 & 0.60 & 0.09 & - \\ 
0.10 & 40.0 & 6.28 & 0.32 & 0.61 & 0.07 & - \\ 
0.10 & 50.0 & 5.02 & 0.38 & 0.59 & 0.03 & - \\ 
0.34 & 20.0 & 7.11 & 0.28 & 0.56 & 0.15 & - \\ 
0.34 & 25.0 & 6.98 & 0.29 & 0.57 & 0.14 & - \\ 
0.34 & 27.5 & 6.86 & 0.29 & 0.57 & 0.13 & - \\ 
0.34 & 30.0 & 6.69 & 0.30 & 0.57 & 0.13 & - \\ 
0.34 & 35.0 & 6.38 & 0.31 & 0.57 & 0.11 & - \\ 
0.34 & 40.0 & 6.07 & 0.33 & 0.57 & 0.10 & - \\ 
0.34 & 45.0 & 5.40 & 0.32 & 0.59 & 0.09 & 0.28 \\ 
0.34 & 50.0 & 5.09 & 0.33 & 0.59 & 0.08 & 0.20 \\ 
0.50 & 20.0 & 7.12 & 0.28 & 0.56 & 0.16 & - \\ 
0.50 & 25.0 & 6.90 & 0.29 & 0.56 & 0.15 & - \\ 
0.50 & 27.5 & 6.72 & 0.30 & 0.56 & 0.14 & - \\ 
0.50 & 30.0 & 6.44 & 0.30 & 0.56 & 0.14 & 0.20 \\ 
0.50 & 35.0 & 5.99 & 0.30 & 0.57 & 0.14 & 0.18 \\ 
0.50 & 40.0 & 5.80 & 0.30 & 0.58 & 0.13 & 0.18 \\ 
0.50 & 45.0 & 5.62 & 0.31 & 0.58 & 0.12 & 0.14 \\ 
0.50 & 50.0 & 5.43 & 0.31 & 0.58 & 0.10 & 0.10 \\ 
0.71 & 20.0 & 7.12 & 0.28 & 0.56 & 0.16 & - \\ 
0.71 & 25.0 & 6.28 & 0.27 & 0.56 & 0.17 & 0.79 \\ 
0.71 & 27.5 & 6.20 & 0.27 & 0.56 & 0.17 & 0.80 \\ 
0.71 & 30.0 & 6.13 & 0.28 & 0.56 & 0.16 & 0.80 \\ 
0.71 & 35.0 & 5.99 & 0.28 & 0.57 & 0.15 & 0.81 \\ 
0.71 & 40.0 & 5.91 & 0.28 & 0.57 & 0.15 & 0.78 \\ 
0.71 & 45.0 & 5.84 & 0.29 & 0.58 & 0.14 & 0.77 \\ 
0.71 & 50.0 & 5.77 & 0.29 & 0.58 & 0.13 & 0.73 \\ 
0.94 & 20.0 & 7.00 & 0.29 & 0.56 & 0.15 & - \\ 
0.94 & 25.0 & 6.18 & 0.27 & 0.55 & 0.18 & 0.94 \\ 
0.94 & 27.5 & 6.13 & 0.27 & 0.55 & 0.18 & 0.95 \\ 
0.94 & 30.0 & 6.10 & 0.27 & 0.55 & 0.17 & 0.96 \\ 
0.94 & 35.0 & 6.04 & 0.28 & 0.56 & 0.17 & 0.96 \\ 
0.94 & 40.0 & 6.00 & 0.28 & 0.56 & 0.16 & 0.97 \\ 
0.94 & 45.0 & 5.96 & 0.28 & 0.56 & 0.16 & 0.96 \\ 
0.94 & 50.0 & 5.92 & 0.28 & 0.57 & 0.15 & 0.96 \\ 
1.10 & 20.0 & 6.65 & 0.27 & 0.55 & 0.18 & 0.54 \\ 
1.10 & 25.0 & 6.16 & 0.27 & 0.55 & 0.18 & 0.98 \\ 
1.10 & 27.5 & 6.13 & 0.27 & 0.55 & 0.18 & 1.00 \\ 
1.10 & 30.0 & 6.11 & 0.27 & 0.55 & 0.18 & 0.99 \\ 
1.10 & 35.0 & 6.09 & 0.27 & 0.55 & 0.18 & 1.00 \\ 
1.10 & 40.0 & 6.06 & 0.27 & 0.55 & 0.17 & 0.99 \\ 
1.10 & 45.0 & 6.03 & 0.28 & 0.55 & 0.17 & 0.99 \\ 
1.10 & 50.0 & 6.00 & 0.28 & 0.56 & 0.17 & 0.99 \\ 
1.30 & 20.0 & 6.23 & 0.27 & 0.54 & 0.19 & 0.98 \\ 
1.30 & 25.0 & 6.19 & 0.27 & 0.54 & 0.19 & 1.00 \\ 
1.30 & 27.5 & 6.18 & 0.27 & 0.54 & 0.19 & 1.00 \\ 
1.30 & 30.0 & 6.16 & 0.27 & 0.54 & 0.19 & 1.00 \\ 
1.30 & 35.0 & 6.15 & 0.27 & 0.54 & 0.18 & 1.00 \\ 
1.30 & 40.0 & 6.13 & 0.27 & 0.55 & 0.18 & 1.00 \\ 
1.30 & 45.0 & 6.11 & 0.27 & 0.55 & 0.18 & 1.00 \\ 
1.30 & 50.0 & 6.08 & 0.27 & 0.55 & 0.18 & 1.00 \\ 
\hline
\end{tabular}
\end{table}

\begin{table}
\caption{A summary of the results of the simulations of collisions of the $2\,\mathrm{M_\oplus}$ projectile with the $6.25\,\mathrm{M_\oplus}$ target. Full planet data is given in Table \ref{tab:PlanetData}, the rest of the collision results are in Tables \ref{tab:CollisionData_1}, \ref{tab:CollisionData_3}, \ref{tab:CollisionData_4}, and \ref{tab:CollisionData_5}.}
\label{tab:CollisionData_2}
\begin{tabular}{cccccccccc}
\hline
$b$ & $v$ & $M_{\mathrm{LR}}$ & $f_{\mathrm{Fe}}$ & $f_{\mathrm{Si}}$ & $f_{\mathrm{H}}$ & $M_{\mathrm{SLR}}$ \\ 
 & $\mathrm{km\,s^{-1}}$ & $\mathrm{M_{\oplus}}$ & & & & $\mathrm{M_{\oplus}}$\\ 
\hline
0.00 & 20.0 & 7.94 & 0.29 & 0.59 & 0.12 & - \\ 
0.00 & 25.0 & 7.72 & 0.30 & 0.60 & 0.09 & - \\ 
0.00 & 30.0 & 7.51 & 0.31 & 0.62 & 0.07 & - \\ 
0.00 & 35.0 & 6.93 & 0.34 & 0.62 & 0.05 & - \\ 
0.00 & 40.0 & 6.07 & 0.38 & 0.58 & 0.04 & - \\ 
0.00 & 45.0 & 5.09 & 0.44 & 0.55 & 0.02 & - \\ 
0.00 & 50.0 & 3.82 & 0.48 & 0.52 & - & - \\ 
0.10 & 20.0 & 7.98 & 0.29 & 0.59 & 0.12 & - \\ 
0.10 & 25.0 & 7.77 & 0.30 & 0.60 & 0.10 & - \\ 
0.10 & 27.5 & 7.69 & 0.30 & 0.61 & 0.09 & - \\ 
0.10 & 30.0 & 7.52 & 0.31 & 0.61 & 0.07 & - \\ 
0.10 & 35.0 & 7.05 & 0.33 & 0.61 & 0.06 & - \\ 
0.10 & 40.0 & 6.20 & 0.38 & 0.59 & 0.04 & - \\ 
0.10 & 50.0 & 3.85 & 0.48 & 0.51 & 0.01 & - \\ 
0.34 & 20.0 & 8.04 & 0.29 & 0.58 & 0.13 & - \\ 
0.34 & 25.0 & 7.85 & 0.30 & 0.59 & 0.11 & - \\ 
0.34 & 27.5 & 7.54 & 0.31 & 0.59 & 0.10 & 0.14 \\ 
0.34 & 30.0 & 7.25 & 0.32 & 0.58 & 0.09 & - \\ 
0.34 & 35.0 & 5.75 & 0.31 & 0.60 & 0.09 & 1.12 \\ 
0.34 & 40.0 & 5.36 & 0.32 & 0.60 & 0.08 & 0.96 \\ 
0.34 & 45.0 & 4.98 & 0.34 & 0.60 & 0.06 & 0.81 \\ 
0.34 & 50.0 & 4.61 & 0.36 & 0.59 & 0.05 & 0.56 \\ 
0.50 & 20.0 & 8.07 & 0.29 & 0.58 & 0.13 & - \\ 
0.50 & 25.0 & 7.85 & 0.30 & 0.59 & 0.11 & - \\ 
0.50 & 27.5 & 7.81 & 0.30 & 0.59 & 0.11 & - \\ 
0.50 & 30.0 & 6.08 & 0.30 & 0.57 & 0.13 & 1.64 \\ 
0.50 & 35.0 & 5.77 & 0.30 & 0.58 & 0.12 & 1.45 \\ 
0.50 & 40.0 & 5.57 & 0.30 & 0.59 & 0.11 & 1.35 \\ 
0.50 & 50.0 & 5.15 & 0.32 & 0.60 & 0.08 & 1.04 \\ 
0.71 & 20.0 & 8.11 & 0.29 & 0.58 & 0.14 & - \\ 
0.71 & 25.0 & 6.19 & 0.27 & 0.57 & 0.16 & 1.85 \\ 
0.71 & 27.5 & 6.13 & 0.27 & 0.57 & 0.16 & 1.82 \\ 
0.71 & 30.0 & 6.06 & 0.28 & 0.57 & 0.15 & 1.82 \\ 
0.71 & 35.0 & 5.91 & 0.28 & 0.58 & 0.14 & 1.79 \\ 
0.71 & 40.0 & 5.78 & 0.29 & 0.58 & 0.13 & 1.77 \\ 
0.71 & 50.0 & 5.63 & 0.30 & 0.59 & 0.12 & 1.72 \\ 
0.94 & 20.0 & 7.94 & 0.29 & 0.57 & 0.13 & - \\ 
0.94 & 25.0 & 6.14 & 0.27 & 0.55 & 0.18 & 1.96 \\ 
0.94 & 27.5 & 6.10 & 0.27 & 0.55 & 0.17 & 1.97 \\ 
0.94 & 30.0 & 6.06 & 0.27 & 0.55 & 0.17 & 1.97 \\ 
0.94 & 35.0 & 6.01 & 0.28 & 0.56 & 0.17 & 1.98 \\ 
0.94 & 40.0 & 5.95 & 0.28 & 0.56 & 0.16 & 1.97 \\ 
0.94 & 50.0 & 5.86 & 0.28 & 0.57 & 0.15 & 1.98 \\ 
1.10 & 20.0 & 8.06 & 0.29 & 0.56 & 0.15 & 0.13 \\ 
1.10 & 25.0 & 6.14 & 0.27 & 0.55 & 0.18 & 2.00 \\ 
1.10 & 27.5 & 6.12 & 0.27 & 0.55 & 0.18 & 2.01 \\ 
1.10 & 30.0 & 6.10 & 0.27 & 0.55 & 0.18 & 2.01 \\ 
1.10 & 35.0 & 6.06 & 0.27 & 0.55 & 0.17 & 2.00 \\ 
1.10 & 40.0 & 6.04 & 0.28 & 0.55 & 0.17 & 2.00 \\ 
1.10 & 50.0 & 5.98 & 0.28 & 0.56 & 0.16 & 2.00 \\ 
1.30 & 20.0 & 6.20 & 0.27 & 0.54 & 0.19 & 2.01 \\ 
1.30 & 25.0 & 6.18 & 0.27 & 0.54 & 0.19 & 2.01 \\ 
1.30 & 27.5 & 6.17 & 0.27 & 0.54 & 0.19 & 2.00 \\ 
1.30 & 30.0 & 6.16 & 0.27 & 0.54 & 0.19 & 2.01 \\ 
1.30 & 35.0 & 6.15 & 0.27 & 0.54 & 0.19 & 2.00 \\ 
1.30 & 40.0 & 6.12 & 0.27 & 0.55 & 0.18 & 2.00 \\ 
1.30 & 50.0 & 6.08 & 0.27 & 0.55 & 0.18 & 2.00 \\ 
\hline
\end{tabular}
\end{table}

\begin{table}
\caption{A summary of the results of the simulations of collisions of the $3\,\mathrm{M_\oplus}$ projectile with the $6.25\,\mathrm{M_\oplus}$ target. Full planet data is given in Table \ref{tab:PlanetData}, the rest of the collision results are in Tables \ref{tab:CollisionData_1}, \ref{tab:CollisionData_2}, \ref{tab:CollisionData_4}, and \ref{tab:CollisionData_5}.}
\label{tab:CollisionData_3}
\begin{tabular}{cccccccccc}
\hline
$b$ & $v$ & $M_{\mathrm{LR}}$ & $f_{\mathrm{Fe}}$ & $f_{\mathrm{Si}}$ & $f_{\mathrm{H}}$ & $M_{\mathrm{SLR}}$ \\ 
 & $\mathrm{km\,s^{-1}}$ & $\mathrm{M_{\oplus}}$ & & & & $\mathrm{M_{\oplus}}$\\ 
\hline
0.00 & 20.0 & 8.75 & 0.31 & 0.60 & 0.09 & - \\ 
0.00 & 25.0 & 8.72 & 0.31 & 0.61 & 0.08 & - \\ 
0.00 & 30.0 & 8.41 & 0.32 & 0.62 & 0.06 & - \\ 
0.00 & 35.0 & 7.45 & 0.36 & 0.60 & 0.04 & - \\ 
0.00 & 40.0 & 6.39 & 0.41 & 0.56 & 0.03 & - \\ 
0.00 & 45.0 & 5.01 & 0.44 & 0.54 & 0.01 & - \\ 
0.00 & 50.0 & 3.51 & 0.52 & 0.48 & - & - \\ 
0.10 & 20.0 & 8.86 & 0.30 & 0.60 & 0.10 & - \\ 
0.10 & 25.0 & 8.69 & 0.31 & 0.61 & 0.08 & - \\ 
0.10 & 27.5 & 8.60 & 0.31 & 0.62 & 0.07 & - \\ 
0.10 & 30.0 & 8.41 & 0.32 & 0.62 & 0.06 & - \\ 
0.10 & 35.0 & 7.64 & 0.35 & 0.60 & 0.05 & - \\ 
0.10 & 40.0 & 6.53 & 0.41 & 0.56 & 0.03 & - \\ 
0.10 & 50.0 & 3.21 & 0.52 & 0.48 & - & - \\ 
0.34 & 20.0 & 9.01 & 0.30 & 0.59 & 0.11 & - \\ 
0.34 & 25.0 & 8.76 & 0.31 & 0.61 & 0.09 & - \\ 
0.34 & 27.5 & 8.60 & 0.31 & 0.61 & 0.08 & - \\ 
0.34 & 30.0 & 8.24 & 0.32 & 0.62 & 0.05 & 0.07 \\ 
0.34 & 35.0 & 5.62 & 0.31 & 0.61 & 0.08 & 2.08 \\ 
0.34 & 40.0 & 5.19 & 0.33 & 0.61 & 0.06 & 1.78 \\ 
0.34 & 45.0 & 4.76 & 0.35 & 0.60 & 0.05 & 1.57 \\ 
0.34 & 50.0 & 4.24 & 0.39 & 0.58 & 0.03 & 1.31 \\ 
0.50 & 20.0 & 9.04 & 0.30 & 0.59 & 0.11 & - \\ 
0.50 & 25.0 & 8.79 & 0.30 & 0.61 & 0.09 & - \\ 
0.50 & 27.5 & 8.82 & 0.30 & 0.60 & 0.09 & - \\ 
0.50 & 30.0 & 5.91 & 0.29 & 0.59 & 0.12 & 2.82 \\ 
0.50 & 35.0 & 5.64 & 0.30 & 0.59 & 0.11 & 2.67 \\ 
0.50 & 40.0 & 5.35 & 0.31 & 0.59 & 0.09 & 2.47 \\ 
0.50 & 45.0 & 5.20 & 0.32 & 0.60 & 0.08 & 2.29 \\ 
0.50 & 50.0 & 4.98 & 0.33 & 0.60 & 0.07 & 2.09 \\ 
0.71 & 20.0 & 9.11 & 0.29 & 0.59 & 0.12 & - \\ 
0.71 & 25.0 & 6.11 & 0.27 & 0.57 & 0.15 & 2.91 \\ 
0.71 & 27.5 & 6.05 & 0.28 & 0.58 & 0.15 & 2.88 \\ 
0.71 & 30.0 & 5.98 & 0.28 & 0.58 & 0.14 & 2.85 \\ 
0.71 & 35.0 & 5.83 & 0.29 & 0.58 & 0.14 & 2.81 \\ 
0.71 & 40.0 & 5.69 & 0.29 & 0.58 & 0.13 & 2.80 \\ 
0.71 & 45.0 & 5.62 & 0.30 & 0.58 & 0.12 & 2.75 \\ 
0.71 & 50.0 & 5.55 & 0.30 & 0.59 & 0.11 & 2.74 \\ 
0.94 & 20.0 & 9.11 & 0.29 & 0.58 & 0.12 & - \\ 
0.94 & 25.0 & 6.12 & 0.27 & 0.55 & 0.17 & 2.99 \\ 
0.94 & 27.5 & 6.06 & 0.27 & 0.56 & 0.17 & 3.00 \\ 
0.94 & 30.0 & 6.03 & 0.28 & 0.56 & 0.17 & 2.99 \\ 
0.94 & 35.0 & 5.98 & 0.28 & 0.56 & 0.16 & 2.99 \\ 
0.94 & 40.0 & 5.93 & 0.28 & 0.56 & 0.15 & 2.99 \\ 
0.94 & 45.0 & 5.88 & 0.28 & 0.57 & 0.15 & 2.99 \\ 
0.94 & 50.0 & 5.83 & 0.29 & 0.57 & 0.14 & 2.98 \\ 
1.10 & 20.0 & 9.15 & 0.29 & 0.58 & 0.13 & - \\ 
1.10 & 25.0 & 6.12 & 0.27 & 0.55 & 0.18 & 3.02 \\ 
1.10 & 27.5 & 6.11 & 0.27 & 0.55 & 0.18 & 3.02 \\ 
1.10 & 30.0 & 6.10 & 0.27 & 0.55 & 0.18 & 3.01 \\ 
1.10 & 35.0 & 6.06 & 0.27 & 0.55 & 0.17 & 3.01 \\ 
1.10 & 40.0 & 6.03 & 0.28 & 0.56 & 0.17 & 3.01 \\ 
1.10 & 45.0 & 5.99 & 0.28 & 0.56 & 0.16 & 3.00 \\ 
1.10 & 50.0 & 5.97 & 0.28 & 0.56 & 0.16 & 3.00 \\ 
1.30 & 20.0 & 6.19 & 0.27 & 0.54 & 0.19 & 3.03 \\ 
1.30 & 25.0 & 6.18 & 0.27 & 0.54 & 0.19 & 3.01 \\ 
1.30 & 27.5 & 6.18 & 0.27 & 0.54 & 0.19 & 3.01 \\ 
1.30 & 30.0 & 6.16 & 0.27 & 0.54 & 0.19 & 3.01 \\ 
1.30 & 35.0 & 6.14 & 0.27 & 0.55 & 0.18 & 3.01 \\ 
1.30 & 40.0 & 6.12 & 0.27 & 0.55 & 0.18 & 3.01 \\ 
1.30 & 45.0 & 6.10 & 0.27 & 0.55 & 0.18 & 3.01 \\ 
1.30 & 50.0 & 6.08 & 0.27 & 0.55 & 0.18 & 3.01 \\ 
\hline
\end{tabular}
\end{table}

\begin{table}
\caption{A summary of the results of the simulations of collisions of the $4\,\mathrm{M_\oplus}$ projectile with the $6.25\,\mathrm{M_\oplus}$ target. Full planet data is given in Table \ref{tab:PlanetData}, the rest of the collision results are in Tables \ref{tab:CollisionData_1}, \ref{tab:CollisionData_2}, \ref{tab:CollisionData_3}, and \ref{tab:CollisionData_5}.}
\label{tab:CollisionData_4}
\begin{tabular}{cccccccccc}
\hline
$b$ & $v$ & $M_{\mathrm{LR}}$ & $f_{\mathrm{Fe}}$ & $f_{\mathrm{Si}}$ & $f_{\mathrm{H}}$ & $M_{\mathrm{SLR}}$ \\ 
 & $\mathrm{km\,s^{-1}}$ & $\mathrm{M_{\oplus}}$ & & & & $\mathrm{M_{\oplus}}$\\ 
\hline
0.00 & 20.0 & 9.73 & 0.31 & 0.62 & 0.08 & - \\ 
0.00 & 25.0 & 9.69 & 0.31 & 0.62 & 0.07 & - \\ 
0.00 & 30.0 & 9.40 & 0.32 & 0.63 & 0.05 & - \\ 
0.00 & 35.0 & 8.32 & 0.36 & 0.60 & 0.04 & - \\ 
0.00 & 40.0 & 7.11 & 0.42 & 0.56 & 0.03 & - \\ 
0.00 & 45.0 & 5.45 & 0.44 & 0.55 & 0.01 & - \\ 
0.00 & 50.0 & 3.70 & 0.53 & 0.47 & - & - \\ 
0.10 & 20.0 & 9.83 & 0.30 & 0.61 & 0.09 & - \\ 
0.10 & 25.0 & 9.62 & 0.31 & 0.62 & 0.07 & - \\ 
0.10 & 27.5 & 9.51 & 0.31 & 0.63 & 0.06 & - \\ 
0.10 & 30.0 & 9.41 & 0.32 & 0.63 & 0.05 & - \\ 
0.10 & 35.0 & 8.42 & 0.35 & 0.61 & 0.04 & - \\ 
0.10 & 40.0 & 7.20 & 0.41 & 0.57 & 0.02 & - \\ 
0.10 & 50.0 & 3.08 & 0.53 & 0.47 & - & 0.13 \\ 
0.34 & 20.0 & 9.96 & 0.30 & 0.60 & 0.10 & - \\ 
0.34 & 25.0 & 9.68 & 0.31 & 0.62 & 0.07 & - \\ 
0.34 & 27.5 & 9.52 & 0.31 & 0.63 & 0.06 & - \\ 
0.34 & 30.0 & 9.27 & 0.32 & 0.64 & 0.04 & - \\ 
0.34 & 35.0 & 5.50 & 0.31 & 0.62 & 0.07 & 3.62 \\ 
0.34 & 40.0 & 5.05 & 0.33 & 0.61 & 0.05 & 2.91 \\ 
0.34 & 45.0 & 4.49 & 0.37 & 0.60 & 0.04 & 2.44 \\ 
0.34 & 50.0 & 3.86 & 0.42 & 0.56 & 0.02 & 2.13 \\ 
0.50 & 20.0 & 10.01 & 0.30 & 0.60 & 0.10 & - \\ 
0.50 & 25.0 & 9.74 & 0.31 & 0.62 & 0.08 & - \\ 
0.50 & 27.5 & 9.45 & 0.32 & 0.63 & 0.06 & - \\ 
0.50 & 30.0 & 5.75 & 0.30 & 0.60 & 0.11 & 3.94 \\ 
0.50 & 35.0 & 5.56 & 0.30 & 0.60 & 0.10 & 3.73 \\ 
0.50 & 40.0 & 5.28 & 0.32 & 0.60 & 0.08 & 3.51 \\ 
0.50 & 45.0 & 5.01 & 0.33 & 0.60 & 0.07 & 3.31 \\ 
0.50 & 50.0 & 4.83 & 0.34 & 0.60 & 0.06 & 3.11 \\ 
0.71 & 20.0 & 10.09 & 0.30 & 0.60 & 0.11 & - \\ 
0.71 & 25.0 & 9.97 & 0.30 & 0.60 & 0.10 & - \\ 
0.71 & 27.5 & 5.95 & 0.28 & 0.58 & 0.14 & 3.96 \\ 
0.71 & 30.0 & 5.89 & 0.28 & 0.58 & 0.14 & 3.93 \\ 
0.71 & 35.0 & 5.75 & 0.29 & 0.58 & 0.13 & 3.86 \\ 
0.71 & 40.0 & 5.62 & 0.30 & 0.58 & 0.12 & 3.82 \\ 
0.71 & 45.0 & 5.52 & 0.30 & 0.59 & 0.11 & 3.78 \\ 
0.71 & 50.0 & 5.48 & 0.30 & 0.59 & 0.10 & 3.74 \\ 
0.94 & 20.0 & 10.16 & 0.29 & 0.59 & 0.11 & - \\ 
0.94 & 25.0 & 6.08 & 0.27 & 0.56 & 0.17 & 4.02 \\ 
0.94 & 27.5 & 6.04 & 0.28 & 0.56 & 0.17 & 4.02 \\ 
0.94 & 30.0 & 6.01 & 0.28 & 0.56 & 0.16 & 4.01 \\ 
0.94 & 35.0 & 5.95 & 0.28 & 0.56 & 0.16 & 4.00 \\ 
0.94 & 40.0 & 5.90 & 0.28 & 0.57 & 0.15 & 4.00 \\ 
0.94 & 45.0 & 5.85 & 0.28 & 0.57 & 0.14 & 4.00 \\ 
0.94 & 50.0 & 5.81 & 0.29 & 0.57 & 0.14 & 4.00 \\ 
1.10 & 20.0 & 10.07 & 0.30 & 0.60 & 0.11 & - \\ 
1.10 & 25.0 & 6.12 & 0.27 & 0.55 & 0.18 & 4.03 \\ 
1.10 & 27.5 & 6.10 & 0.27 & 0.55 & 0.18 & 4.03 \\ 
1.10 & 30.0 & 6.09 & 0.27 & 0.55 & 0.18 & 4.02 \\ 
1.10 & 35.0 & 6.06 & 0.27 & 0.55 & 0.17 & 4.01 \\ 
1.10 & 40.0 & 6.02 & 0.28 & 0.56 & 0.17 & 4.01 \\ 
1.10 & 45.0 & 5.98 & 0.28 & 0.56 & 0.16 & 4.01 \\ 
1.10 & 50.0 & 5.95 & 0.28 & 0.56 & 0.16 & 4.01 \\ 
1.30 & 20.0 & 6.18 & 0.27 & 0.54 & 0.19 & 4.04 \\ 
1.30 & 25.0 & 6.18 & 0.27 & 0.54 & 0.19 & 4.02 \\ 
1.30 & 27.5 & 6.17 & 0.27 & 0.54 & 0.19 & 4.01 \\ 
1.30 & 30.0 & 6.17 & 0.27 & 0.54 & 0.19 & 4.01 \\ 
1.30 & 40.0 & 6.11 & 0.27 & 0.55 & 0.18 & 4.01 \\ 
1.30 & 45.0 & 6.09 & 0.27 & 0.55 & 0.18 & 4.01 \\ 
\hline
\end{tabular}
\end{table}

\begin{table}
\caption{A summary of the results of the simulations of collisions of the $5\,\mathrm{M_\oplus}$ projectile with the $6.25\,\mathrm{M_\oplus}$ target. Full planet data is given in Table \ref{tab:PlanetData}, the rest of the collision results are in Tables \ref{tab:CollisionData_1}, \ref{tab:CollisionData_2}, \ref{tab:CollisionData_3}, and \ref{tab:CollisionData_4}.}
\label{tab:CollisionData_5}
\begin{tabular}{cccccccccc}
\hline
$b$ & $v$ & $M_{\mathrm{LR}}$ & $f_{\mathrm{Fe}}$ & $f_{\mathrm{Si}}$ & $f_{\mathrm{H}}$ & $M_{\mathrm{SLR}}$ \\ 
 & $\mathrm{km\,s^{-1}}$ & $\mathrm{M_{\oplus}}$ & & & & $\mathrm{M_{\oplus}}$\\ 
\hline
0.00 & 20.0 & 10.63 & 0.31 & 0.62 & 0.06 & - \\ 
0.00 & 25.0 & 10.66 & 0.31 & 0.62 & 0.06 & - \\ 
0.00 & 30.0 & 10.40 & 0.32 & 0.63 & 0.05 & - \\ 
0.00 & 35.0 & 9.35 & 0.36 & 0.61 & 0.03 & - \\ 
0.00 & 40.0 & 7.99 & 0.42 & 0.56 & 0.02 & - \\ 
0.00 & 45.0 & 6.25 & 0.44 & 0.55 & 0.01 & - \\ 
0.00 & 50.0 & 4.45 & 0.52 & 0.48 & - & - \\ 
0.10 & 20.0 & 10.81 & 0.31 & 0.62 & 0.08 & - \\ 
0.10 & 25.0 & 10.56 & 0.32 & 0.63 & 0.06 & - \\ 
0.10 & 27.5 & 10.47 & 0.32 & 0.63 & 0.05 & - \\ 
0.10 & 30.0 & 10.40 & 0.32 & 0.63 & 0.05 & - \\ 
0.10 & 35.0 & 9.48 & 0.35 & 0.61 & 0.03 & - \\ 
0.10 & 40.0 & 8.08 & 0.41 & 0.57 & 0.02 & - \\ 
0.10 & 50.0 & 4.09 & 0.51 & 0.49 & - & 0.01 \\ 
0.34 & 20.0 & 10.93 & 0.30 & 0.61 & 0.09 & - \\ 
0.34 & 25.0 & 10.67 & 0.31 & 0.62 & 0.07 & - \\ 
0.34 & 27.5 & 10.61 & 0.31 & 0.63 & 0.06 & - \\ 
0.34 & 30.0 & 10.29 & 0.32 & 0.64 & 0.04 & - \\ 
0.34 & 35.0 & 5.36 & 0.31 & 0.63 & 0.06 & 4.86 \\ 
0.34 & 40.0 & 4.88 & 0.34 & 0.62 & 0.04 & 4.36 \\ 
0.34 & 45.0 & 4.24 & 0.38 & 0.59 & 0.03 & 3.71 \\ 
0.34 & 50.0 & 3.52 & 0.45 & 0.53 & 0.02 & 3.15 \\ 
0.50 & 20.0 & 11.03 & 0.30 & 0.61 & 0.09 & - \\ 
0.50 & 25.0 & 10.72 & 0.31 & 0.62 & 0.07 & - \\ 
0.50 & 27.5 & 10.43 & 0.32 & 0.63 & 0.05 & - \\ 
0.50 & 30.0 & 5.67 & 0.29 & 0.61 & 0.10 & 4.98 \\ 
0.50 & 35.0 & 5.47 & 0.30 & 0.61 & 0.09 & 4.83 \\ 
0.50 & 40.0 & 5.16 & 0.32 & 0.60 & 0.07 & 4.63 \\ 
0.50 & 45.0 & 4.90 & 0.34 & 0.60 & 0.06 & 4.39 \\ 
0.50 & 50.0 & 4.64 & 0.35 & 0.60 & 0.05 & 4.20 \\ 
0.71 & 20.0 & 11.08 & 0.30 & 0.60 & 0.10 & - \\ 
0.71 & 25.0 & 10.95 & 0.30 & 0.61 & 0.09 & - \\ 
0.71 & 27.5 & 5.90 & 0.28 & 0.58 & 0.14 & 5.00 \\ 
0.71 & 30.0 & 5.81 & 0.29 & 0.58 & 0.13 & 4.99 \\ 
0.71 & 35.0 & 5.66 & 0.29 & 0.58 & 0.12 & 4.93 \\ 
0.71 & 40.0 & 5.53 & 0.30 & 0.58 & 0.12 & 4.87 \\ 
0.71 & 45.0 & 5.49 & 0.30 & 0.59 & 0.11 & 4.83 \\ 
0.71 & 50.0 & 5.42 & 0.31 & 0.59 & 0.10 & 4.75 \\ 
0.94 & 20.0 & 11.15 & 0.30 & 0.60 & 0.10 & - \\ 
0.94 & 25.0 & 6.04 & 0.28 & 0.56 & 0.17 & 5.05 \\ 
0.94 & 27.5 & 6.01 & 0.28 & 0.56 & 0.16 & 5.04 \\ 
0.94 & 30.0 & 5.97 & 0.28 & 0.56 & 0.16 & 5.03 \\ 
0.94 & 35.0 & 5.92 & 0.28 & 0.56 & 0.15 & 5.01 \\ 
0.94 & 40.0 & 5.88 & 0.28 & 0.57 & 0.15 & 5.01 \\ 
0.94 & 45.0 & 5.83 & 0.29 & 0.57 & 0.14 & 5.00 \\ 
0.94 & 50.0 & 5.79 & 0.29 & 0.58 & 0.14 & 5.00 \\ 
1.10 & 20.0 & 11.09 & 0.30 & 0.60 & 0.10 & - \\ 
1.10 & 25.0 & 6.10 & 0.27 & 0.55 & 0.18 & 5.05 \\ 
1.10 & 27.5 & 6.09 & 0.27 & 0.55 & 0.18 & 5.03 \\ 
1.10 & 30.0 & 6.08 & 0.27 & 0.55 & 0.18 & 5.03 \\ 
1.10 & 35.0 & 6.05 & 0.28 & 0.55 & 0.17 & 5.02 \\ 
1.10 & 40.0 & 6.02 & 0.28 & 0.56 & 0.17 & 5.01 \\ 
1.10 & 45.0 & 5.98 & 0.28 & 0.56 & 0.16 & 5.01 \\ 
1.10 & 50.0 & 5.94 & 0.28 & 0.56 & 0.16 & 5.01 \\ 
1.30 & 20.0 & 6.17 & 0.27 & 0.54 & 0.19 & 5.05 \\ 
1.30 & 25.0 & 6.17 & 0.27 & 0.54 & 0.19 & 5.02 \\ 
1.30 & 27.5 & 6.17 & 0.27 & 0.54 & 0.19 & 5.02 \\ 
1.30 & 30.0 & 6.17 & 0.27 & 0.54 & 0.19 & 5.02 \\ 
1.30 & 35.0 & 6.15 & 0.27 & 0.54 & 0.19 & 5.01 \\ 
1.30 & 40.0 & 6.12 & 0.27 & 0.55 & 0.18 & 5.01 \\ 
1.30 & 45.0 & 6.09 & 0.27 & 0.55 & 0.18 & 5.01 \\ 
1.30 & 50.0 & 6.06 & 0.27 & 0.55 & 0.17 & 5.01 \\ 
\hline
\end{tabular}
\end{table}

\section{A Modified Prescription For The Mass Of the Largest Remnant}
\label{ap:Prescription}

Using our new results we can now add to the collision prescription from \citet{Denman2020} to account for different impact angles. It should be noted that we have not yet tested the effects of target mass and target atmosphere mass on oblique collisions.  
The new prescription is as follows:

\begin{enumerate}
    \item For a given collision scenario with a $6.25\,\mathrm{M_{\oplus}}$ target with $20\%$ atmosphere by mass ($M_\mathrm{p}$, $v$ and $b$), first calculate the specific kinetic energy of the impact: 
    \begin{equation}
        Q_{\mathrm{R}}=\frac{1}{2}\mu\frac{v^2}{M_{\mathrm{tot}}}.
    \end{equation}
    \item Then calculate the transition velocity between erosive hit-and-run and accretion/disruption events, $v_{\mathrm{split}}$, which is given by 
    \begin{equation}
        v_{\mathrm{split}}=\sqrt{\frac{2G(M_{\mathrm{t}}+M_{\mathrm{p}})}{b(R_{\mathrm{t}}+R_{\mathrm{p}})}}.
    \end{equation}
    \item If $v<v_{\mathrm{split}}$ then follow the prescription from \citet{Denman2020}, outlined here, otherwise skip to \ref{it:ehr}. 
    \begin{enumerate}
        \item Calculate the specific kinetic energy of the transition between the atmosphere loss and core and mantle loss regimes,
        \begin{equation}
            Q_{\mathrm{piv}}=\left(\frac{M_{\mathrm{tot}}}{M_\oplus}\right)\left(-2.45\frac{M_\mathrm{p}}{M_\mathrm{t}}+14.56\right) \;\; [10^6\,\mathrm{J\,kg^{-1}}].
        \end{equation}
    
        \item  Calculate the catastrophic disruption threshold,
        \begin{equation}
        Q_{\mathrm{RD}}^{*}=c^*\frac{4}{5}\pi\rho_{\mathrm{1}}GR_{\mathrm{C1}}^2,
        \end{equation}
        where $\rho_{\mathrm{1}}=1000\,\mathrm{kg\,m^{-3}}$, and $R_{\mathrm{C1}} = \left(\frac{3M_{\mathrm{tot}}}{4\pi\rho_{\mathrm{1}}}\right)^{\frac{1}{3}}$ is the radius of a spherical body with a density of $\rho_{\mathrm{1}}$ and mass $M_{\mathrm{tot}}$. We use the value $c^*=2.52$ predicted by \citet{Denman2020} to determine the catastrophic disruption threshold ($Q_{\mathrm{RD}}^{*(\mathrm{New})}$), as well as the value $c^*=1.9$ from \citet{Leinhardt2012} for atmosphere-less planets which is used to predict energy dependence in the core and mantle dominated loss regime ($Q_{\mathrm{RD}}^{*(\mathrm{LS12})}$). 
        \\

        \item Then, calculate the specific energy gradient for largest remnant mass for the core and mantle loss regime:
        \begin{equation}
            m_{\mathrm{c\&m}}=\frac{-0.5(1+f_{\mathrm{H}}^{\mathrm{t}})}{Q_{\mathrm{RD}}^{*(\mathrm{LS12})}}.
        \end{equation} 
        where $f_{\mathrm{H}}^{\mathrm{t}}$ is the mass fraction of the target which is atmosphere.

        \item Next, calculate the gradient for the atmosphere loss dominated regime from the core and mantle loss gradient assuming zero impact energy means zero mass loss,
        \begin{equation}
            m_{\mathrm{atmos}}=\frac{m_{\mathrm{c\&m}}(Q_{\mathrm{piv}}-Q_{\mathrm{RD}}^{*(\mathrm{New})})-0.5}{Q_{\mathrm{piv}}}.
        \end{equation}

        \item The next step is to calculate the super-catastrophic disruption threshold ($M_{\mathrm{LR}}<0.1M_{\mathrm{tot}}$), we use the \citet{Leinhardt2012} method for this,
        \begin{equation}      
        Q_{\mathrm{supercat}}=Q_{\mathrm{RD}}^{*(\mathrm{New})}-\frac{0.4}{m_{\mathrm{c\&m}}}.
        \end{equation}
    
        \item From the above three steps the total mass in the largest remnant must therefore be:
        \begin{equation}
            \frac{M_{\mathrm{LR}}}{M_{\mathrm{tot}}}=
            \begin{cases}
             m_{\mathrm{atmos}}Q_{\mathrm{R}}+1 & 0<Q_{\mathrm{R}}<Q_{\mathrm{piv}} \\ \\
             m_{\mathrm{c\&m}}(Q_{\mathrm{R}}-Q_{\mathrm{RD}}^{*(\mathrm{New})})+0.5 & Q_{\mathrm{piv}}<Q_{\mathrm{R}}<Q_{\mathrm{supercat}}.\\
             
            \end{cases}.
        \end{equation}
        As neither this study nor \citet{Denman2020} probed the super-catastrophic disruption regime we do not predict results here, we recommend the \citet{Leinhardt2012} prescription for collisions in this regime.\\

        \item  Finally, the mass of atmosphere in the largest remnant is given by:
        \begin{equation}
        \frac{M^{\mathrm{H}}_{\mathrm{LR}}}{M^{\mathrm{H}}_{\mathrm{t}}}
        =
        \begin{cases} 
          1+\frac{0.94^2}{4}\left(\frac{Q_{\mathrm{R}}}{Q_{\mathrm{piv}}}\right)^{2}-0.94\frac{Q_{\mathrm{R}}}{Q_{\mathrm{piv}}} & \frac{Q_{\mathrm{R}}}{Q_{\mathrm{piv}}}<2.12 \\
          0 & \frac{Q_{\mathrm{R}}}{Q_{\mathrm{piv}}}>2.12.
        \end{cases}
        \end{equation}
    \end{enumerate}

    \item \label{it:ehr} If, on the other hand, $v>v_{\mathrm{split}} $ then we have an erosive hit-and-run collision and the following process should be used to predict largest remnant mass (note we have not tested the dependencies of these power laws on target mass or target atmosphere fraction):
    \begin{enumerate}
        \item The total atmosphere mass remaining in the largest remnant is given by
        \begin{equation}            \frac{M^{\mathrm{H}}_{\mathrm{LR}}}{M^{\mathrm{H}}_{\mathrm{t}}}
            =
           -10^{-0.76b+0.18}\log_{10}(Q_\mathrm{R})+10^{-0.54b+1.02},
        \end{equation}
        \item and the mass remaining in the largest remnant is
        \begin{equation}
            M_{\mathrm{LR}}= -10^{-1.11b-8.56}Q_\mathrm{R}+M_\mathrm{t}.
        \end{equation}
    \end{enumerate}
 \end{enumerate}


\bsp	
\label{lastpage}
\end{document}